\documentclass[twocolumn, aps, prb, showpacs]{revtex4-1}

\usepackage{amsmath}
\usepackage{amssymb}
\usepackage{bm}
\usepackage{color}
\usepackage{graphicx}
\usepackage[latin1]{inputenc}
\usepackage[english]{babel}
\begin{document}


\title{Bias dependence of perpendicular spin torque and of free and fixed layer eigenmodes in MgO-based nanopillars}
\author{P.~K.~Muduli}
\email{pranaba.muduli@physics.gu.se}
\affiliation{Physics Department, University of Gothenburg, 412 96 Gothenburg, Sweden}

\author{O.~G.~Heinonen}
\affiliation{Materials Science Division, Argonne National Laboratory, Lemont, IL 60439, USA}

\author{Johan~\AA kerman}
\affiliation{Physics Department, University of Gothenburg, 412 96 Gothenburg, Sweden}
\affiliation{Materials Physics, Royal Institute of Technology,
Electrum 229, 164 40 Kista, Sweden}

\begin{abstract}
We have measured the bias voltage and field dependence of eigenmode frequencies in a magnetic tunnel junction with MgO barrier. We show that both free layer (FL) and reference layer (RL) modes are excited,  and that a cross-over between these modes is observed by varying external field and bias voltage. The bias voltage dependence of the FL and RL modes are shown to be dramatically different. The bias dependence of the FL modes is linear in bias voltage, whereas that of the RL mode is strongly quadratic. Using modeling and micromagnetic simulations, we show that the linear bias dependence of  FL frequencies is primarily due to a linear dependence of the perpendicular spin torque on bias voltage, whereas the quadratic dependence of the RL on bias voltage is dominated by the reduction of exchange bias due to Joule heating, and is not attributable to a quadratic dependence of the perpendicular spin torque on bias voltage.
\end{abstract}

 \maketitle

\section{INTRODUCTION}
A current flowing in a magnetic structure will be spin-polarized because of spin-asymmetric scattering in the two spin channels (we shall assume that the spin relaxation length is longer than any relevant system dimension). In non-collinear systems or magnetic heterostructures in which the magnetization in different layers is aligned in different directions, the spin polarized conduction electrons can exert a torque on the magnetization order parameters. This spin transfer torque couples the direct current with magnetization dynamics and allows  for the manipulation of magnetization using spin-polarized currents.~\cite{slonczewski1996jmmm,berger1996prb} In a magnetic multilayer system, the spin transfer torque can be used to pump energy into the magnetization dynamics so as to precisely offset dissipative losses, leading to self-sustaining magnetization oscillations. These so-called spin torque oscillators (STOs) can potentially be used for microwave signal generation,~\cite{tsoi2000nt,kiselev2003nt} modulation,~\cite{pufall2005apl,muduli2010prb} and detection.~\cite{tulapurkar2005nt} Recently, STOs based on CoFeB/MgO/CoFeB magnetic tunnel junctions (MTJs) have attracted considerable interest because of their relatively large microwave power.~\cite{deac2008np,nazarov2008jap,houssameddine2008apl} In addition to the in-plane spin torque predicted by Slonczewski and Berger,~\cite{slonczewski1996jmmm,berger1996prb} the perpendicular spin torque first predicted by Zhang, Levy, and Fert~\cite{zhang2002prl} is appreciable in MTJs~\cite{wang2009prb,li2008prl} and much larger than in metallic systems.~\cite{xia2002prb,zimmler2004prb,urazdin2003prl} As the perpendicular spin torque couples to the magnetization dynamics, it is of significant interest to determine this component for fundamental understanding as well as for future application.~\cite{zhou2009apl}

The perpendicular spin torque in an MTJ depends on the current, and therefore on the bias voltage applied across the MTJ. Several methods have been proposed and used to determine the magnitude and bias dependence of the perpendicular spin torque.~\cite{petit2007prl,sankey2008ntp,kubota2008ntp,deac2008np,li2008prl,wang2009prb,oh2009ntp,heinonen2010prb,heinonen2010prl} The reported results vary both in their qualitative and quantitative assessment of the bias dependence of the perpendicular spin torque. Some studies show a linear,~\cite{petit2007prl,heinonen2010prl}  others show a quadratic,~\cite{sankey2008ntp,kubota2008ntp,wang2009prb,jung2010prb} or a  combination of linear and quadratic~\cite{oh2009ntp} behavior of the perpendicular spin torque with bias voltage. Early theoretical works predicted a quadratic dependence of spin torque on bias voltage.~\cite{theodonis2006prl,stiles2008prb} A recent theory however predict a linear~\cite{xiao2008prb} contribution to perpendicular spin torque for asymmetric MTJs, in which the two magnetic layers on either side of the tunnel barrier are not identical, which was the case for Ref.~\onlinecite{oh2009ntp}. However, in a recent study of some 400 MTJ devices combined with micromagnetic simulations~\cite{heinonen2010prl} it was shown that the perpendicular torque varies linearly with bias voltage (for a certain range of bias voltages) even for nominally symmetric MTJs, and that any quadratic term is negligible.

\begin{figure}[t!]
\includegraphics*[width=.45\textwidth]{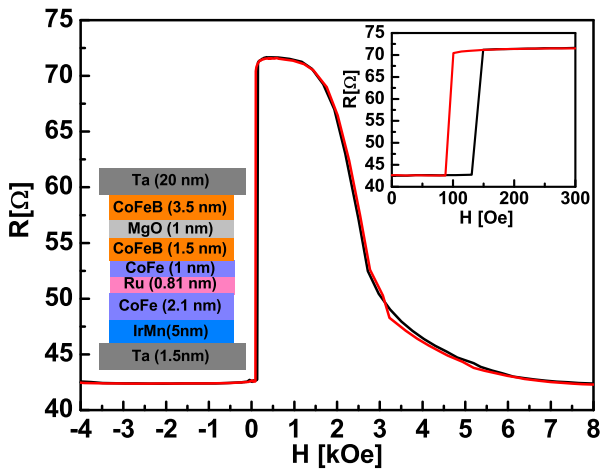}
\caption{(Color online) Magnetoresistance loop of the MTJ nanopillar showing TMR of 70\%. The top inset shows a magnified view of the low field region whereas the bottom inset shows the layer stack of the MTJ.}\label{fig:fig1}
\end{figure}

We will here present results for the bias and field dependence of eigenmode frequencies for  MTJs with MgO tunneling barriers and nominally symmetric interfaces on each side of the tunneling barrier. In reality, there will be some asymmetry introduced by deposition kinetics and thermal anneal, but even large, deliberately introduced asymmetry have been shown to have a relatively small effect on the
perpendicular spin torque\cite{oh2009ntp} and we do not expect that asymmetries play any significant role in the behavior of the
MTJs studied here. A cartoon of the cross-section of the devices is shown in the inset of Fig.~\ref{fig:fig1}. An antiferromagnetic IrMn layer provides an exchange bias on the CoFe pinned layer (PL) which tends to keep its magnetization direction in a fixed direction. The PL is strongly coupled antiferromagnetically through a 0.81 nm thick Ru layer to the composite CoFe/CoFeB fixed layer, or reference layer (RL). This CoFe/Ru/CoFe/CoFeB structure is referred to as a synthetic antiferromagnet (SAF) as its net magnetic moment is close to zero. Above the MgO tunnel barrier is the CoFeB free layer (FL) which can easily be rotated by an external field.  We will show that FL as well as RL magnetization eigenmodes can be observed, depending on the strength of the applied magnetic field. While the observation of magnetization modes in the GHz range has been already reported for similar devices,~\cite{cornelissen2009epl,cornelissen2010prb,houssameddine2010apl} the detailed bias dependence of the FL and RL modes and the role of perpendicular spin torque on these modes have not been discussed. For the first time, we show that the bias dependence of the FL and RL modes are dramatically different. The bias dependence of the FL modes is approximately linear in bias voltage, whereas the RL modes show a strong parabolic dependence with bias voltage. Using modeling and micromagnetic simulations, we show that this linear bias dependence of  the FL frequencies is primarily due to a linear dependence of perpendicular spin torque, whereas the parabolic decrease of frequency of the RL with bias voltage is dominated by the reduction of exchange bias due to Joule heating.

\section{Experiment}
\begin{figure}[t!]
\includegraphics*[width=.45\textwidth]{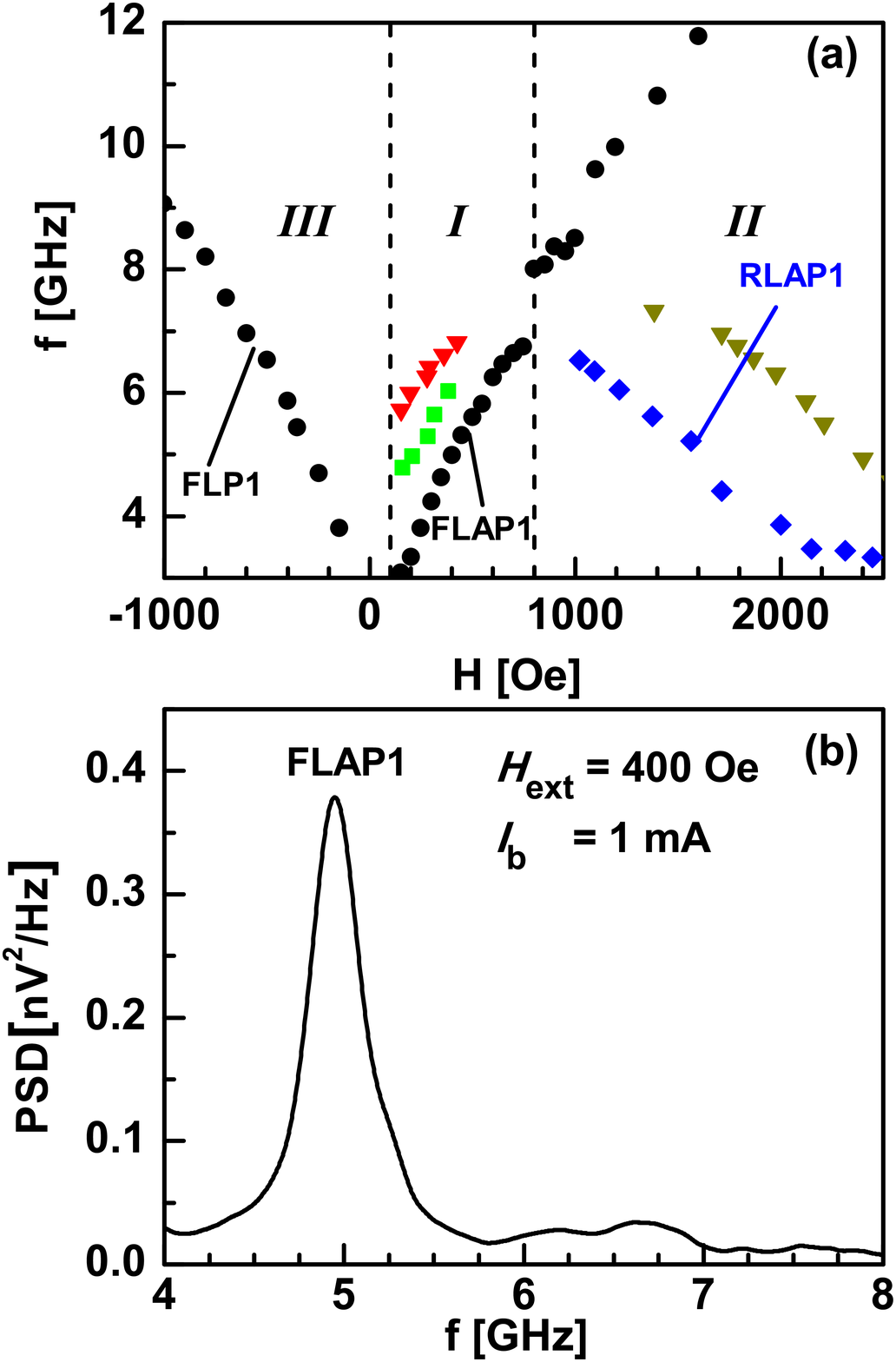}
\caption{(Color online) (a) Mode frequencies versus easy axis external field measured at low bias current of 1~mA. The dashed lines define the boundary of the three regions, \emph{I}, \emph{II} and \emph{III} as described in the text. (b)~Example spectra measured at $H_{\rm ext}=400$~Oe and low bias current of $I_{\rm b}=1$~mA.}\label{fig:fig2}
\end{figure}

The MTJ samples were fabricated using state-of-the-art industrial ultra-high-vacuum sputter deposition.~\cite{mao2006ieeem,nazarov2006apl,nazarov2008jap} The deposition technique typically produces layers with a root mean square roughness of about 0.1~nm after thermal anneal. We will here predominantly discuss room temperature results from a circular device with an approximate diameter of 240~nm. However, similar results are also obtained in a number of other devices of same dimensions, as well as several devices with diameter of 180~nm. The magnetization densities of the FL and RL are approximately 1000 emu/cm$^3$, and that of the pinned layer approximately $1200$~emu/cm$^3$.~\cite{kubota2006apl,huai2006jjap,heinonen2010prl} The direction of the exchange bias in all devices is defined as the $\hat{\mathbf{x}}$-direction. The measurement set-up is similar to that described in Ref.~\onlinecite{bonetti2009apl}. The signal generated from the STO was amplified using a broadband 0.1-26~GHz, +45~dB microwave amplifier, and detected by a spectrum analyzer. The \textit{dc} bias current is fed to the device by a current source through a 0-26~GHz bias-tee connected in parallel with the transmission line. We use the convention that a positive current flows from the FL to the RL, in which case the electrons flow from the RL to the FL.

In Fig.~\ref{fig:fig1}, a magnetoresistance curve is shown where positive field is along the $\hat{\mathbf{x}}$-direction which is also the direction of the exchange bias on the PL. The measured tunneling magnetoresistance (TMR) is about 70\% and the resistance area (RA) product in the parallel state is about 1.5~$\Omega~(\mu$m$)^{2}$. We found the expected strong bias dependence of TMR and decrease of resistance with temperature, both of which confirm the integrity of tunneling barrier and absence of shorts.~\cite{akerman2000apl,rabson2001jap,akerman2001apl,akerman2002jmmm,teixeira2009jap,fuchs2006prl,akerman2003epl} The temperature dependence of the TMR in our devices is measured to be -0.13\%/K. The bias range was limited to $|V_{\rm b}|<0.4$~V to avoid any risk of damage to the MTJs due to prolonged exposure to bias voltages. During the measurement, we continuously measure resistance of the device to make sure that the device is not shorted. After application of high voltages ($|V_{\rm b}|>0.6$~V), some samples exhibited a strong reduction of TMR similar to Ref~\onlinecite{houssameddine2008apl}. Such low TMR samples are not investigated in this work. We observe a loop shift of about 100~Oe of the $R-H$ curve (as shown in the inset) which is a result of RL-FL ferromagnetic interlayer exchange coupling (IEC) due to, e.g., barrier roughness (N{\'e}el orange-peel coupling) or electronic processes such as coherent scattering processes that also give rise to an RKKY-like coupling. The coercive field of the FL is approximately $20$~Oe. At higher fields, the resistance decreases due to rotation of the SAF. The two SAF layers (PL and RL) starts to rotate when the net field on the PL exceeds the exchange bias field. We observe a small hysteresis in this regime indicating that the rotation of the two layers is not fully coherent.~\cite{parkin1999jap,helmer2010prb} From the rotation of the SAF, we define the exchange bias field $H_{\rm eb}$ using the peak in the first derivative of resistance versus field on the positive field side. The magnetoresistance loop of Fig.~\ref{fig:fig1} was measured at a low bias $I_{\rm b}=0.1$~mA, from which we found $H_{\rm eb}=2200$~Oe.

\begin{figure}[t!]
\includegraphics*[width=.45\textwidth]{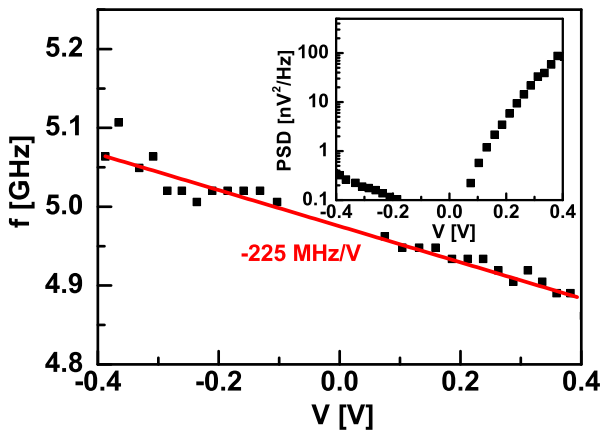}
\caption{(Color online) Bias dependence of mode FLAP1 at $H_{\rm ext}=400$~Oe and a linear fit (red solid line). The inset shows power spectral density (PSD) of this mode as function of bias voltage.}\label{fig:fig3}
\end{figure}

\section{Results}

\begin{figure}[t!]
\includegraphics*[width=.45\textwidth]{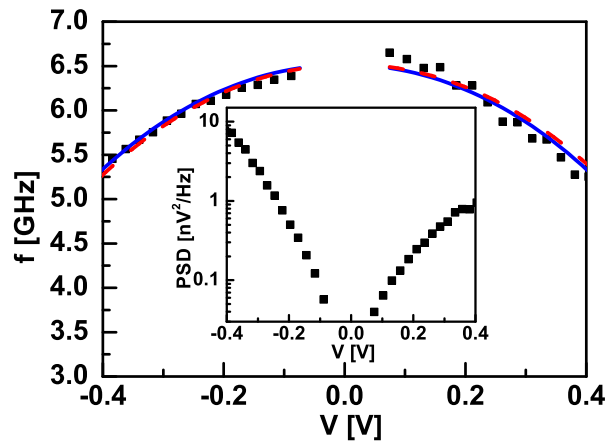}
\caption{(Color online) Measured bias dependence of the mode RLAP1 at $H_{\rm ext}$=1000~Oe (solid squares) and calculations using Eqs.~(\ref{eq:RLAP1_vs_V_b}) and (\ref{eq:H_eb_vs_V_b}) for $b_J=b_1=0$ (blue solid line), and $b_J=b_1V_{\rm b}$ with $b_1=37$~Oe/V (red dashed line). The inset shows power (PSD) of this mode as function of bias voltage.}\label{fig:fig4}
\end{figure}

In Fig.~\ref{fig:fig2}(a), we show the low-bias behavior of frequency versus external field, $H_{\rm ext}$. A bias current of $I_{\rm b}=1$~mA was used for this measurement (which corresponds to about 42~mV in the P state and 73~mV in AP state). The frequency of the modes was determined from the measured spectra. An example spectra is shown in Fig.~\ref{fig:fig2}(b) for $H_{\rm ext}=400$~Oe and $I_{\rm b}=1$~mA. Based on the low bias measurement of Fig.~\ref{fig:fig2}(a), we distinguish three regions in which three different behaviors of the eigenmodes are observed. The first region (\emph{I}) corresponds to positive field, $100<H_{\rm ext}\alt800$~Oe in which the FL and RL magnetizations are in an antiparallel (AP) configuration. In this region the frequencies of the modes increase with external field. As we will demonstrate later, these modes correspond to free layer excitations and we will label them FLAP. The second region (\emph{II}) corresponds to higher positive fields, $H_{\rm ext}\agt800$~Oe, where additional lower-frequency modes appear with frequencies that decrease with $H_{\rm ext}$. We will identify these new modes as RL modes, and we will label them RLAP. Finally, the third region (\emph{III}) corresponds to the case of $H_{\rm ext}<100$~Oe with the magnetization of the FL and RL in parallel (P) configuration. The frequency of the single mode visible in this region increases with external field strength. We will identify this mode too as a FL mode. To distinguish from the FL modes in the AP configuration we will label it as FLP1. To study the bias dependence we identify three modes in these three regions labeled as FLAP1, RLAP1, and FLP1 as shown in Fig.~\ref{fig:fig2}(a), which are strongest (with highest power) modes  in the respective regions.

\begin{figure}[t!]
\includegraphics*[width=.45\textwidth]{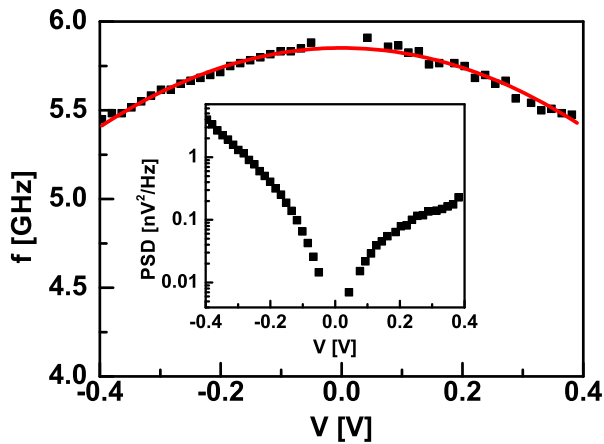}
\caption{(Color online) Bias dependence of mode FLP1 at $H_{\rm ext}=-400$~Oe and a parabolic fit (red solid line). The inset shows power (PSD) of this mode as function of bias voltage.}\label{fig:fig5}
\end{figure}

The bias dependencies of the frequency and power of the FLAP1, RLAP1, and FLP1 modes are shown in Fig.~\ref{fig:fig3}, Fig.~\ref{fig:fig4} and Fig.~\ref{fig:fig5}, respectively. The field values were chosen such that apparent mode-crossings are avoided and there is minimum mixing of modes from different branches. Firstly, the FLAP1 mode shows a linear behavior of frequency with bias voltage for $|V_{\rm b}|\alt0.4$~V. The power spectral density shows a strong polarity-dependent emission power and increases almost exponentially with bias voltage. Since the structure is in the AP state at $H_{\rm ext}=400$~Oe, spin torque tends to destabilize the FL magnetization for positive polarity. Hence higher power for positive polarity is consistent with spin torque excitation of FL layer.

The frequency of the RLAP1 mode on the other hand shows a strong parabolic behavior as a function of bias voltage (Fig.~\ref{fig:fig4}), and a negligible linear contribution. The bias dependence of the frequency is hence symmetric in bias voltage in contrast to that of the FLAP1 mode, which is linear in bias voltage and hence antisymmetric with $V_{\rm b}$ (apart from a trivial constant). The power of the RLAP1 also increases exponentially with bias voltage, but shows the opposite bias asymmetry, i.e. it is higher for electrons flowing from FL to RL, indicating that the excitation is a RL mode. As will be discussed in the next section, this is also confirmed by micromagnetic simulations and modeling.

The frequency of the FLP1 mode in Fig.~\ref{fig:fig5} also shows a parabolic dependence on bias voltage, which is again in contrast to that of the FLAP1 mode. However, the parabolic dependence is much weaker than that of the RLAP1 mode with a curvature that is only about 30\% of the RLAP1 mode's curvature. The power of this mode also increases with bias voltage. However, the maximum power is by about one order of magnitude lower compared to the RLAP1 mode and about two orders of magnitude lower compared to that of the FLAP1 mode. Interestingly, the power has a polarity-dependent asymmetry similar to that of the RLAP1 mode.

\section{Discussion}\label{sec:discussion}

\subsection{Low-bias mode frequencies}\label{sec:discussion_low_bias}

We start by analyzing the low-bias behavior of the FLAP1, FLP1 and RLAP1 modes shown in Fig.~\ref{fig:fig2}(a). We take a coordinate system with $\hat{x}$ along the positive field direction (along the direction of the exchange-bias field on the pinned layer), $\hat y$ perpendicular to the field and in the plane of the magnetic layers, and $\hat z$ perpendicular to the plane of the layers. A simple Kittel-like model can be obtained for the lowest bulk-like FLAP1 mode (we are for now ignoring edge modes) in the following way. We represent the magnetization density as a macro-spin vector ${\bm M}_{\rm S}(t)=M_{\rm S}{\bm m}(t)$, where $M_{\rm S}$ is the saturation magnetization density of the FL and ${\bm m}(t)$ a unit vector along the magnetization direction. If we assume that stray fields from the SAF can be ignored, the effective field acting on the FL in the AP configuration with the RL magnetization along $-\hat x$ is

\begin{widetext}
\begin{equation}
{\bm H}_{\rm eff}=\left(H_{\rm ext}-H_{\rm IEC}\right)\hat x-M_{\rm S}\left(N_x m_x\hat x+N_y m_y\hat y\right)-b_J\hat y-H_dm_z\hat z.
\label{eq:FLAP1_H_eff}
\end{equation}
\end{widetext}
Here $H_{\rm IEC}$ is the effective field due to the RL-FL IEC, $N_x$ and $N_y$ are the
$xx$ and $yy$ components of the demagnetizing tensor (assuming an ellipsoidal shape for which the demagnetizing tensor is diagonal), $H_d=4\pi M_{\rm S}$ is the out-of-plane demagnetizing field, and $b_J$ is the effective field due to perpendicular spin torque. We have here included $b_J$ for completeness, but in the low-bias limit we will later set $b_J=0$.
By linearizing the torque equation
$d{\bm M}_{\rm S}/dt=-|\gamma_e|{\bm M}_{\rm S}\times {\bm H}_{\rm eff}$  in small excursions about the equilibrium magnetization configuration, ${\bm m(t)}=\hat x+\delta{\bm m}(t)$, with $|\delta {\bm m}|\ll 1$ and perpendicular to $\hat x$, we obtain for the zero-bias resonant frequency $\omega_{{\rm FLAP1},0}$
\begin{figure}[t!]
\includegraphics*[width=.45\textwidth]{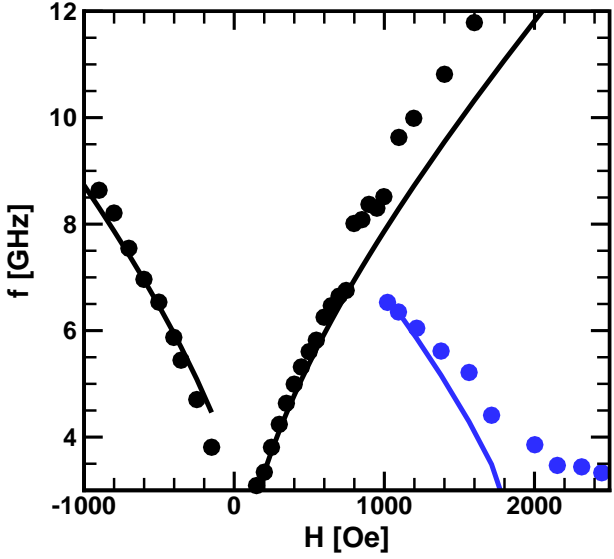}
\caption{(Color online) Fit to the FLAP1, RLAP1, and FLP1 modes of Fig.~\ref{fig:fig2}(a) using Eqs.~(\ref{eq:FLAP1_Kittel}) - (\ref{eq:RLAP1_Kittel})}\label{fig:Kittel_fit_300x600}
\end{figure}
\begin{figure}[t!]
\includegraphics*[width=.47\textwidth]{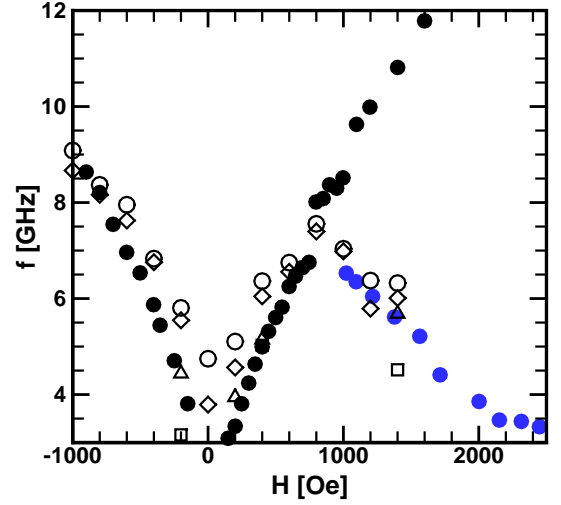}
\caption{Experimental values (filled circles) for the modes FLAP1 (black), RLAP1 (blue), and FLP1 (black), together with micromagnetically
calculated frequencies for the lowest-lying bulk modes (squares, triangles, diamonds, and circles). The agreement is satisfactory, and the micromagnetic results show a cross-over to a branch with decreasing frequency as a function of magnetic field, similar to the experimentally measured
cross-over from the FLAP1 to the RLAP1 modes.}\label{fig:expt_mumag_circle}
\end{figure}
\begin{widetext}
\begin{equation}
\omega_{{\rm FLAP1},0}^2=\gamma_e^2\left[\left( N_y-N_x\right) M_{\rm S}+H_{\rm ext}-H_{\rm IEC}-b_J\right]\left[H_d-N_xM_{\rm S}+H_{\rm ext}-H_{\rm IEC}-b_J\right],
\label{eq:FLAP1_Kittel}
\end{equation}
\end{widetext}
with $\gamma_e$ is the electron gyromagnetic factor, $|\gamma_e|\approx2.8$~GHz/kOe. Equation~\ref{eq:FLAP1_Kittel} lends itself to a simple interpretation: apart from the demagnetizing fields, the main stiffness and restoring torque is provided by the external field, while the ferromagnetic IEC softens the mode, since the RL magnetization is in the $-\hat x$ direction. In the P configuration, we similarly obtain for the zero-bias FLP1 mode (assuming the RL is stationary)
\begin{widetext}
\begin{equation}
\omega_{{\rm FLP1},0}^2=\gamma_e^2 \left[\left( N_y-N_x\right) M_{\rm S}+H_{\rm ext}+H_{\rm IEC}+b_J\right]\left[H_d-N_xM_{\rm S}+H_{\rm ext}+H_{\rm IEC}+b_J\right].
\label{eq:FLP1_Kittel}
\end{equation}

Finally, again under the assumption that we can ignore stray fields from the SAF, we obtain for the zero-bias RLAP1 mode

\begin{eqnarray}
\omega_{{\rm RLAP1},0}^2=\gamma_e^2 \left[N_x M_{\rm S}-H_{\rm ext}+H_{\rm eb}-H_{\rm IEC}+b_J\right] \left[H_d+N_xM_{\rm S}-H_{\rm ext}+H_{\rm eb}-H_{\rm IEC}+b_J \right].
\label{eq:RLAP1_Kittel}
\end{eqnarray}
\end{widetext}

For this mode, the external field softens the mode as it is antiparallel to the equilibrium direction of the RL magnetization, while the IEC stiffens the mode. It is now clear that
Eqs.~(\ref{eq:FLAP1_Kittel}) - (\ref{eq:RLAP1_Kittel}) at least qualitatively can explain the behavior of the FLAP1, RLAP1, and FLP1 modes: The cross-over from FLAP1 to RLAP1 modes occurs when the external field is large enough (for a given IEC) that it drives the RLAP1 frequency below that of the FLAP1 mode. Figure~\ref{fig:Kittel_fit_300x600} shows a fit to the data in Fig.~\ref{fig:fig2}(a) (positive bias) using Eqs.~(\ref{eq:FLAP1_Kittel})-(\ref{eq:RLAP1_Kittel}). For the FLAP1 and FLP1 modes the same parameters were used, with
$(N_y-N_x)M_{\rm S}=100$~Oe, $H_{\rm IEC}=100$~Oe, and $H_d-N_xM_{\rm S}=7$~kOe, and $b_J=0$. To fit RLAP1 mode, we used Eq.~\ref{eq:RLAP1_Kittel} with $N_xM_{\rm S}=-15$~Oe,
$H_{\rm eb}=2.2$~kOe, and $H_d+N_xM_{\rm S}=4.0$~kOe.~\cite{kubota2006apl,huai2006jjap,heinonen2010prl} The values for the FLAP1 and FLP1 modes are quite reasonable, especially because the dimensions of the device have not been determined accurately, but only the target dimensions from the lithographic process are known. Also, the magnetization densities for the layers are estimated based on values from similar devices and process conditions. The values used for the RLAP1 fit are also quite reasonable in view of these facts. This simple analysis supports our identification of the modes as FL modes (FLAP1 and FLP1), and a RL mode (RLAP1), and therefore also the cross-over of the
lowest-lying mode from FLAP1 to RLAP1 at $H_{\rm ext}\approx800$~Oe.

\begin{figure}[t!]
\includegraphics*[width=.45\textwidth]{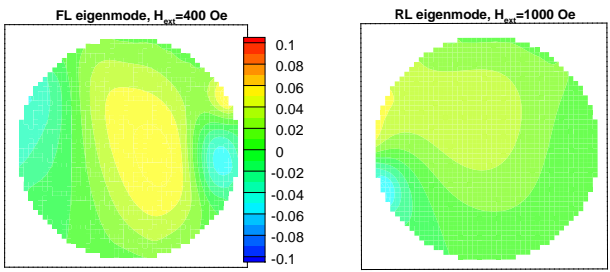}
\caption{(Color online) Eigenmode distributions for the two lowest eigenmodes at $H_{\rm ext}=400$~Oe (left panel) and $H_{\rm ext}=1000$~Oe
(right panel). In the figure, the distribution of
the real part of the $x$-component of the eigenmode is plotted for the FL at 400 Oe and the RL at 1000 Oe. At each of these field values,
the exhibited distribution of the eigenmodes was dominant, with distributions in the RL (FL) negligible at 400 Oe (1000 Oe). The figure
clearly shows a cross-over from a FL bulk-like mode, to a bulk-like RL mode.}\label{fig:240x240FL-RL}
\end{figure}

To further  support this analysis, we also performed micromagnetic modeling of a $240\times240$~nm$^2$ device with layer thickness PL=2 nm, RL=3 nm, and FL=3 nm, and magnetization densities as described above; the IEC coupling was taken as 0.035 erg/cm$^2$, corresponding to $H_{\rm IEC}\approx117$~Oe, the pinning energy to the AFM as -1.2 erg/cm$^2$; the intralayer exchange coupling was taken as $1.3\times10^{-6}$ erg/cm, and the dimensionless damping constant $\alpha=0.1$.~\cite{kubota2006apl,huai2006jjap,heinonen2010prl} The system was divided into 5 nm $\times$ 5 nm cells in the $xy$-plane and the magnetization configurations obtained by integrating the time-dependent Landau-Lifshitz-Gilbert (LLG) equations with demagnetizing fields calculated using fast Fourier transforms in the $xy$-plane and direct sums in the $\hat z$-direction. The system was initialized with random magnetization direction in each cell, then subjected to an external field of 4 kOe along the
$\hat x$-direction and the magnetization equilibrated. The field was then changed to 1.4 kOe and subsequently reduced in steps of 200~Oe to -1~kOe, and the magnetization equilibrated at each field value. At each new field value the magnetization was given a small random perturbation from its previous equilibrated configuration.
At each equilibrated magnetization configuration, the LLG equations were linearized about the equilibrium and the resulting homogeneous linear equations solved for the
(complex) magnetization eigenmodes. Figure~\ref{fig:expt_mumag_circle} shows the experimentally measured frequencies for FLAP1, FLP1, and RLAP1 modes together with the real part of the calculated frequencies for the lowest-lying bulk modes (some modes that were clearly edge modes were excluded from the plot). The qualitative behavior of the calculated frequencies agrees with that of the measured ones, and the
quantitative agreement is reasonable. The main point here is that the calculated frequencies for these branches all decrease with increasing field from about 800 Oe. Detailed examination of the distribution of magnetization of these branches confirm that the excited magnetization motion shifts from the FL at 400 Oe to the RL at 1000 Oe (Fig.~\ref{fig:240x240FL-RL}). This cross-over from FL to RL mode is rather insensitive to the parameters used and is also observed for different geometries, {\em e.g.\/,}
$300\times 600$~nm$^2$, $150\times600$~nm$^2$ or $150\times300$~nm$^2$. We also note that the FLP1 branch of these three modes (negative external field) tends to be distributed spatially near an edge on the $\hat{\mathbf{x}}$-axis, although the precise distributions
depend on the particular parameters used. This is not inconsistent with the experimental observation that the power of the FLP1 mode is much lower than that of the FLAP1 and RLAP1 modes.

\subsection{Bias-voltage dependence of the mode frequencies}

We now turn to the bias voltage dependence of the frequencies of the FLAP1, RLAP1, and FLP1 modes. We can safely ignore any bias dependence of magnetization density, IEC, and intralayer exchange, and will focus on the bias dependence of the perpendicular spin torque effective field $b_J$ and the exchange bias field $H_{\rm eb}$. Also, we note that
at these field values the FL and RL remain in P or AP configuration for relatively small bias voltages under consideration, so we take the resistance for each
configuration to be constant and independent of bias voltage. Under the assumption
that the change in $b_J$ and $H_{\rm eb}$ with $V_{\rm b}$ are small compared to their zero-bias values, we can expand $b_J$ and $H_{\rm eb}$ in Taylor series about
$V_{\rm b}=0$. We will ignore any observable direct spin torque effect on the exchange bias.\cite{tsoi2007prl} It is well established that the bias dependence
of $H_{\rm eb}$ decreases monotonically with temperature (and has to vanish at the N\'eel temperature of the AFM).~\cite{nogues2005pr,ali2003prb} For our samples, we measure the temperature dependence to be is -1.65~Oe/K, which is consistent with literature for a similar device.~\cite{georges2009prb} Therefore, the main bias voltage dependence of $H_{\rm eb}$ comes from Joule heating, and $H_{\rm eb}$ must then be an even function of $V_{\rm b}$,
\begin{equation} H_{\rm eb}(V_{\rm b})=H_{\rm eb}^0-aV_{\rm b}^2,
\label{eq:H_eb_vs_V_b}
\end{equation}
where $H_{\rm eb}^0$ is the zero-bias value of $H_{\rm eb}$ and $a$ is a constant of dimension OeV$^{-2}$. In contrast, we will for now expand the perpendicular spin torque
in both linear and quadratic terms,
\begin{equation}
b_J(V_{\rm b})=b_J^0+b_1V_{\rm b}+b_2V_{\rm b}^2,
\label{eq:b_J_vs_V_b}
\end{equation}
\begin{figure}[t!]
\includegraphics*[width=.45\textwidth]{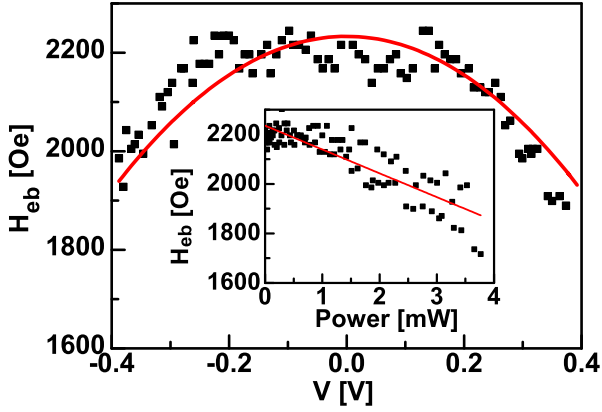}
\caption{(Color online) Bias dependence of $H_{\rm eb}$ and a parabolic fit (red solid line). The inset shows a linear fit of
$H_{\rm eb}$ to applied power.}\label{fig:fig6}
\end{figure}

where $b_J^0$ is the zero-bias value of $b_J$, and the constants $b_1$ and $b_2$ have the dimensions of OeV$^{-1}$ and OeV$^{-2}$, respectively. We will assume that the constants are the same for both AP and P magnetization configurations\cite{oh2009ntp}. Inserting
Eqs.~(\ref{eq:H_eb_vs_V_b}) and (\ref{eq:b_J_vs_V_b}) into Eqs.~(\ref{eq:FLAP1_Kittel})-(\ref{eq:RLAP1_Kittel}) and expanding the square roots to second order in $V_b$ we obtain
\begin{widetext}
\begin{eqnarray}
\label{eq:FLAP1_vs_V_b}
\omega_{\rm FLAP1}(V_{\rm b}) & = & \omega_{{\rm FLAP1},0}
\left\{ 1-
{
\gamma_e^2
\left[ (N_y-2N_x)M_{\rm S}+2H_{\rm ext}-2H_{\rm IEC}+H_d\right]
\over 2\omega_{{\rm FLAP1},0}^2
}
\left[ b_1V_{\rm b}+b_2V_{\rm b}^2\right]
\right\}\\
\label{eq:FLP1_vs_V_b}
\omega_{\rm FLP1}(V_{\rm b}) &=& \omega_{{\rm FLP1},0} \left\{ 1+{
 \gamma_e^2\left[ (N_y-2N_x)M_{\rm S}+2H_{\rm ext}+2H_{\rm IEC}+H_d\right]
\over 2\omega_{{\rm FLP1},0}^2 }
\left[ b_1V_{\rm b}+b_2V_{\rm b}^2\right] \right\}\\
\label{eq:RLAP1_vs_V_b}
\omega_{\rm RLAP1}(V_{\rm b}) & = & \omega_{{\rm RLAP1},0} 
\left\{ 1+
{\gamma_e^2\left[ 2H_{\rm eb}^0+H_d-2N_xM_{\rm S}-2H_{\rm ext}-2H_{\rm IEC}\right]
\over 2\omega_{{\rm RLAP1},0}^2 }
\left[ b_1V_{\rm b}+(b_2-a)V_{\rm b}^2\right] +{\gamma_e^2\over \omega_{{\rm RLAP1},0}^2}b_1^2V_{\rm b}^2\right\}
\end{eqnarray}
\end{widetext}

We first discuss the bias dependence of the FLAP1 mode. If we compare Eq.~(\ref{eq:FLAP1_vs_V_b}) with Fig.~\ref{fig:fig3}, it is clear that the linear dependence
of the frequency of this mode on $V_{\rm b}$ implies that $b_2\approx0$ so that the perpendicular spin torque effective field is linear in $V_{\rm b}$. This is in agreement  with
the results of Petit {\em et al.,}~\cite{petit2007prl} and Heinonen {\em et al.}~\cite{heinonen2010prl}. Using the values for $(Ny-Nx)M_{\rm S}=100$~Oe, $H_{\rm IEC}=100$~Oe, $H_{\rm ext}=400$~Oe, and $H_d-N_xM_{\rm S}=7$~kOe from above, we can estimate the constant
$b_1$ as $b_1\approx37$~Oe/V. With $RA=1.5$~$\Omega\mu$m$^2$ we then obtain a perpendicular spin torque field that is about
$0.55\times 10^{-6}$~Oe(cm)$^2$/A. This is the same order of magnitude as the value of $10^{-6}$~Oe(cm)$^2$/A obtained by
Petit {\em et al.}\cite{petit2007prl} on Al$_2$O$_3$ MTJs with a TMR of 20\%, but an order of magnitude smaller than the value of about $11\times 10^{-6}$~Oe(cm)$^2$/A obtained
by Heinonen {\em et al.}\cite{heinonen2010prl} on MgO MTJs with $RA\approx0.5~\Omega\mu$m$^2$ and a TMR of about 85\%. It is not clear why there is a relatively large spread in these observed values. Tight-binding~\cite{chshiev2008ieeem} and s-d models~\cite{manchon2008jpcm} suggest that the spin torque is sensitive to the particular parameters used in the model ({\em e.g.\/.,} band offsets, s-d coupling) as well as to how much of the transverse spin density is absorbed in the FL. We speculate that the observed large variation in the perpendicular spin torque is a reflection of detailed differences in the MTJs used here and in Refs.~\onlinecite{petit2007prl} and \onlinecite{heinonen2010prl}.

Next, we turn to the bias dependence of the RLAP1 mode, Eq.~(\ref{eq:RLAP1_vs_V_b}). It is clear from this equation that a reduction in $H_{\rm eb}$ due to Joule
heating can have a dominant effect on the bias dependence of this mode's frequency. In order to clarify this effect, we measured $H_{\rm eb}$ as a function of $V_{\rm b}$ (Fig.~\ref{fig:fig6}), from which we obtain $H_{\rm eb}=H_{\rm eb}^0-aV_{\rm b}^2$, with $a=1880$~Oe/V$^2$. We insert this with the {\em same} values as those used in Fig.~\ref{fig:Kittel_fit_300x600} together with $b_1=$37~Oe/V in Eq.~(\ref{eq:RLAP1_vs_V_b})
to get the result shown in Fig.~\ref{fig:fig4}. The scale of the exchange bias field completely dominates the voltage dependence of this mode's frequency compared
to the effect of the spin torque. In Fig.~\ref{fig:fig4} is also exhibited the frequencies obtained by setting $b_1=0$. The result is almost indistinguishable from
that obtained by including the perpendicular spin torque field. We therefore conclude that the parabolic dependence of this mode's frequency on bias voltage
is entirely explained by the weakening of the exchange bias field due to Joule heating, and is not due to spin torque.

In the analysis of the FLAP1 and RLAP1 modes, we have ignored the fact that the resistance in the AP state decreases linearly with increasing magnitude of bias
voltage.~\cite{Dimitrov2009jap} In the present case, the AP resistance
decreases at a rate of about 30~$\Omega$/V. For the FLAP1 mode, we could use current instead of bias voltage as independent variable and consider the change in frequency from its zero-bias value as a function of current. This gives a slope of about -17 GHz/A, which is equal to the slope as a function of bias voltage, -0.23 GHz/V, multiplied
by the zero-bias resistance, 74~$\Omega$, to within experimental uncertainties. Similarly, we can correct for the change in resistance in the analysis of the RLAP1 mode. However,
such a correction does not alter our findings in any qualitative or quantitatively significant way.

Finally, we turn to the bias dependence of the FLP1 mode, which shares many of the characteristics of the RLAP1 mode. As we noted earlier, the bias dependence of this mode's frequency is parabolic, similar to the RLAP1 mode, however with a weaker curvature. The power of the FLP1 mode also shows the same bias voltage asymmetry as the RLAP1 mode, i.e. it is stronger for electrons flowing from the FL to the RL, which is somewhat surprising for an FL mode. It also exhibits a much lower power than either of the RLAP1 or FLAP1 modes. Micromagnetic modeling suggests that this mode is a FL mode, but the precise spatial distribution (edge {\em vs.} bulk) depends rather sensitively on the
specific parameters used in the modeling. To lowest order, the FL modes are not sensitive to the exchange bias strength, so one possibility is that the parabolic bias dependence observed for this mode is indeed due to perpendicular spin torque. A fit to the experimental data in Fig.~\ref{fig:fig5} gives
$f-f_0\approx-2.8$~GHz/V$^2$, which inserted in Eq.~(\ref{eq:FLP1_vs_V_b}) gives $b_2\approx -3\times 10^2$ Oe/V$^2$. This is of opposite sign compared to the values obtained by Oh {\em et al.}\cite{oh2009ntp}, and also larger than their values of 72 and 130 Oe/V$^2$. On the other hand, in the P configuration the resistance of the device is 1.7 times lower than in the AP configuration, so
for the same applied bias voltage, the power dissipated by Joule heating is a factor of 2.8 higher in the P configuration than in the AP one. The Joule heating is not enough to
explain the softening of a FL mode due to reduced magnetization density, which would in turn reduce the restoring torque provided by the
demagnetizing fields $(N_y-N_x)M_{\rm S}$. The Curie temperature of the FL is close to 1000~K, and the temperature necessary to significantly reduce the
magnetization density would have destroyed the tunnel junction. However, if the FLP1 is actually a coupled mode in which both FL and RL oscillate, but with different amplitudes, this can both explain the lower peak power spectral density of this mode as well as the bias dependence of the mode. Any in-phase correlated motion between the FL and the RL will necessarily reduce the resistance fluctuations of the mode, since they are proportional to the cosine of the angle between the RL and FL magnetization directions, and therefore
also the power spectral density of the mode. A main restoring torque of such a coupled RL-FL mode will be provided by the exchange bias, and a reduction in exchange bias
due to Joule heating will soften the mode and can possibly lead to the observed (weak) parabolic dependence of this mode's frequency.

\section{Summary and conclusions}

In summary, we have measured the resonance frequency of a nominally circular MgO tunnel junction with approximately symmetric electrodes, both as a function of applied external magnetic field and as a function of
bias voltage, and analyzed the possible dependence of the perpendicular spin torque effective field as a function of bias voltage for different, fixed magnetic fields. At low magnitudes of magnetic field and low bias voltage, the dominant low-frequency mode (FLAP1) is a FL bulk-like mode. At larger fields in the AP orientation, there is a crossing of this FL branch and a RL branch, and the low-frequency mode is a bulk-like RL mode (RLAP1). At negative fields in the P orientation, the mode appears to be a FL mode (FLP1), that could be more localized to an edge than the FLAP1 and RLAP1 modes. The frequency of the FLAP1 mode at fixed magnetic field has a bias voltage dependence that is linear in bias voltage, which
leads us to conclude that the perpendicular spin torque effective field is linear in bias voltage with a value of about 37 Oe/V, or  $0.55\times 10^{-6}$~Oe(cm)$^2$/A. This is in qualitative agreement with previously observed linear dependence of the perpendicular spin torque,~\cite{heinonen2010prl,petit2007prl} although the magnitude of the effect differs, but in disagreement with the quadratic dependence observed by other groups.~\cite{sankey2008ntp,kubota2008ntp,wang2009prb,jung2010prb} The quadratic behavior of the frequency of the RLAP1 mode that we observed is entirely due to Joule heating, which reduces the exchange bias and consequently the restoring torque for this mode. Any (weak) linear or quadratic dependence on the perpendicular spin torque effective field cannot be extracted beyond the experimental uncertainties. For the FLP1 mode, a much weaker quadratic dependence of the resonant frequency on bias
voltage was observed. We do not believe this quadratic dependence is due to a quadratic dependence of the perpendicular spin torque effective field upon bias voltage. The reason for this is that the sign is opposite to that obtained by Oh {\em et al}~\cite{oh2009ntp} and the effective field would be about a factor of three larger in magnitude. We speculate that this bias dependence of the mode frequency is due to a weak dependence on exchange bias of what is actually a (weakly) coupled FL-RL mode.

\section*{ACKNOWLEDGEMENTS}
Support from the Swedish Foundation for Strategic Research (SSF),
the Swedish Research Council (VR), the G\"{o}ran Gustafsson Foundation and
the Knut and Alice Wallenberg Foundation are gratefully acknowledged. J.~\AA. is a Royal Swedish Academy of Sciences Research Fellow supported by a grant from the Knut and Alice Wallenberg Foundation. O.H. thanks Dr. K. Rivkin for help in setting up the micromagnetic eigenmode calculations, and also acknowledges support from Argonne National Laboratory. Argonne National Laboratory is operated under Contract No. DE-AC02-06CH11357 by UChicago Argonne, LLC.


\begin{thebibliography}{50}%
\makeatletter
\providecommand \@ifxundefined [1]{%
 \@ifx{#1\undefined}
}%
\providecommand \@ifnum [1]{%
 \ifnum #1\expandafter \@firstoftwo
 \else \expandafter \@secondoftwo
 \fi
}%
\providecommand \@ifx [1]{%
 \ifx #1\expandafter \@firstoftwo
 \else \expandafter \@secondoftwo
 \fi
}%
\providecommand \natexlab [1]{#1}%
\providecommand \enquote  [1]{``#1''}%
\providecommand \bibnamefont  [1]{#1}%
\providecommand \bibfnamefont [1]{#1}%
\providecommand \citenamefont [1]{#1}%
\providecommand \href@noop [0]{\@secondoftwo}%
\providecommand \href [0]{\begingroup \@sanitize@url \@href}%
\providecommand \@href[1]{\@@startlink{#1}\@@href}%
\providecommand \@@href[1]{\endgroup#1\@@endlink}%
\providecommand \@sanitize@url [0]{\catcode `\\12\catcode `\$12\catcode
  `\&12\catcode `\#12\catcode `\^12\catcode `\_12\catcode `\%12\relax}%
\providecommand \@@startlink[1]{}%
\providecommand \@@endlink[0]{}%
\providecommand \url  [0]{\begingroup\@sanitize@url \@url }%
\providecommand \@url [1]{\endgroup\@href {#1}{\urlprefix }}%
\providecommand \urlprefix  [0]{URL }%
\providecommand \Eprint [0]{\href }%
\@ifxundefined \urlstyle {%
  \providecommand \doi  [0]{\begingroup \@sanitize@url \@doi}%
  \providecommand \@doi [1]{\endgroup \@@startlink {\doibase
  #1}doi:\discretionary {}{}{}#1\@@endlink }%
}{%
  \providecommand \doi  [0]{doi:\discretionary{}{}{}\begingroup
  \urlstyle{rm}\Url }%
}%
\providecommand \doibase [0]{http://dx.doi.org/}%
\providecommand \Doi [0]{\begingroup \@sanitize@url \@Doi }%
\providecommand \@Doi  [1]{\endgroup\@@startlink{\doibase#1}\@@Doi}%
\providecommand \@@Doi [1]{#1\@@endlink}%
\providecommand \selectlanguage [0]{\@gobble}%
\providecommand \bibinfo  [0]{\@secondoftwo}%
\providecommand \bibfield  [0]{\@secondoftwo}%
\providecommand \translation [1]{[#1]}%
\providecommand \BibitemOpen [0]{}%
\providecommand \bibitemStop [0]{}%
\providecommand \bibitemNoStop [0]{.\EOS\space}%
\providecommand \EOS [0]{\spacefactor3000\relax}%
\providecommand \BibitemShut  [1]{\csname bibitem#1\endcsname}%
\bibitem [{\citenamefont {{Slonczewski}}(1996)}]{slonczewski1996jmmm}%
  \BibitemOpen
  \bibfield  {author} {\bibinfo {author} {\bibfnamefont {J.~C.}\ \bibnamefont
  {{Slonczewski}}},\ }\Doi {10.1016/0304-8853(96)00062-5} {\bibfield  {journal}
  {\bibinfo  {journal} {J.\ Magn.\ Magn.\ Mater.},\ }\textbf {\bibinfo {volume}
  {159}},\ \bibinfo {pages} {L1} (\bibinfo {year} {1996})}\BibitemShut
  {NoStop}%
\bibitem [{\citenamefont {{Berger}}(1996)}]{berger1996prb}%
  \BibitemOpen
  \bibfield  {author} {\bibinfo {author} {\bibfnamefont {L.}~\bibnamefont
  {{Berger}}},\ }\Doi {10.1103/PhysRevB.54.9353} {\bibfield  {journal}
  {\bibinfo  {journal} {\prb},\ }\textbf {\bibinfo {volume} {54}},\ \bibinfo
  {pages} {9353} (\bibinfo {year} {1996})}\BibitemShut {NoStop}%
\bibitem [{\citenamefont {{Tsoi}}\ \emph {et~al.}(2000)\citenamefont {{Tsoi}},
  \citenamefont {{Jansen}}, \citenamefont {{Bass}}, \citenamefont {{Chiang}},
  \citenamefont {{Tsoi}},\ and\ \citenamefont {{Wyder}}}]{tsoi2000nt}%
  \BibitemOpen
  \bibfield  {author} {\bibinfo {author} {\bibfnamefont {M.}~\bibnamefont
  {{Tsoi}}}, \bibinfo {author} {\bibfnamefont {A.~G.~M.}\ \bibnamefont
  {{Jansen}}}, \bibinfo {author} {\bibfnamefont {J.}~\bibnamefont {{Bass}}},
  \bibinfo {author} {\bibfnamefont {W.-C.}\ \bibnamefont {{Chiang}}}, \bibinfo
  {author} {\bibfnamefont {V.}~\bibnamefont {{Tsoi}}}, \ and\ \bibinfo {author}
  {\bibfnamefont {P.}~\bibnamefont {{Wyder}}},\ }\href@noop {} {\bibfield
  {journal} {\bibinfo  {journal} {\nat},\ }\textbf {\bibinfo {volume} {406}},\
  \bibinfo {pages} {46} (\bibinfo {year} {2000})}\BibitemShut {NoStop}%
\bibitem [{\citenamefont {{Kiselev}}\ \emph {et~al.}(2009)\citenamefont
  {{Kiselev}}, \citenamefont {{Sankey}}, \citenamefont {{Krivorotov}},
  \citenamefont {{Emley}}, \citenamefont {{Schoelkopf}}, \citenamefont
  {{Buhrman}},\ and\ \citenamefont {{Ralph}}}]{kiselev2003nt}%
  \BibitemOpen
  \bibfield  {author} {\bibinfo {author} {\bibfnamefont {S.~I.}\ \bibnamefont
  {{Kiselev}}}, \bibinfo {author} {\bibfnamefont {J.~C.}\ \bibnamefont
  {{Sankey}}}, \bibinfo {author} {\bibfnamefont {I.~N.}\ \bibnamefont
  {{Krivorotov}}}, \bibinfo {author} {\bibfnamefont {N.~C.}\ \bibnamefont
  {{Emley}}}, \bibinfo {author} {\bibfnamefont {R.~J.}\ \bibnamefont
  {{Schoelkopf}}}, \bibinfo {author} {\bibfnamefont {R.~A.}\ \bibnamefont
  {{Buhrman}}}, \ and\ \bibinfo {author} {\bibfnamefont {D.~C.}\ \bibnamefont
  {{Ralph}}},\ }\Doi {10.1038/nature01967} {\bibfield  {journal} {\bibinfo
  {journal} {Nature},\ }\textbf {\bibinfo {volume} {425}},\ \bibinfo {pages}
  {380} (\bibinfo {year} {2009})}\BibitemShut {NoStop}%
\bibitem [{\citenamefont {{Pufall}}\ \emph {et~al.}(2005)\citenamefont
  {{Pufall}}, \citenamefont {{Rippard}}, \citenamefont {{Kaka}}, \citenamefont
  {{Silva}},\ and\ \citenamefont {{Russek}}}]{pufall2005apl}%
  \BibitemOpen
  \bibfield  {author} {\bibinfo {author} {\bibfnamefont {M.~R.}\ \bibnamefont
  {{Pufall}}}, \bibinfo {author} {\bibfnamefont {W.~H.}\ \bibnamefont
  {{Rippard}}}, \bibinfo {author} {\bibfnamefont {S.}~\bibnamefont {{Kaka}}},
  \bibinfo {author} {\bibfnamefont {T.~J.}\ \bibnamefont {{Silva}}}, \ and\
  \bibinfo {author} {\bibfnamefont {S.~E.}\ \bibnamefont {{Russek}}},\ }\Doi
  {10.1063/1.1875762} {\bibfield  {journal} {\bibinfo  {journal} {Appl.\ Phys.\
  Lett.},\ }\textbf {\bibinfo {volume} {86}},\ \bibinfo {pages} {082506}
  (\bibinfo {year} {2005})}\BibitemShut {NoStop}%
\bibitem [{\citenamefont {{Muduli}}\ \emph {et~al.}(2010)\citenamefont
  {{Muduli}}, \citenamefont {{Pogoryelov}}, \citenamefont {{Bonetti}},
  \citenamefont {{Consolo}}, \citenamefont {{Mancoff}},\ and\ \citenamefont
  {{{\AA}kerman}}}]{muduli2010prb}%
  \BibitemOpen
  \bibfield  {author} {\bibinfo {author} {\bibfnamefont {P.~K.}\ \bibnamefont
  {{Muduli}}}, \bibinfo {author} {\bibfnamefont {Y.}~\bibnamefont
  {{Pogoryelov}}}, \bibinfo {author} {\bibfnamefont {S.}~\bibnamefont
  {{Bonetti}}}, \bibinfo {author} {\bibfnamefont {G.}~\bibnamefont
  {{Consolo}}}, \bibinfo {author} {\bibfnamefont {F.}~\bibnamefont
  {{Mancoff}}}, \ and\ \bibinfo {author} {\bibfnamefont {J.}~\bibnamefont
  {{{\AA}kerman}}},\ }\Doi {10.1103/PhysRevB.81.140408} {\bibfield  {journal}
  {\bibinfo  {journal} {Phys.\ Rev.\ B},\ }\textbf {\bibinfo {volume} {81}},\
  \bibinfo {pages} {140408} (\bibinfo {year} {2010})}\BibitemShut {NoStop}%
\bibitem [{\citenamefont {{Tulapurkar}}\ \emph {et~al.}(2005)\citenamefont
  {{Tulapurkar}}, \citenamefont {{Suzuki}}, \citenamefont {{Fukushima}},
  \citenamefont {{Kubota}}, \citenamefont {{Maehara}}, \citenamefont
  {{Tsunekawa}}, \citenamefont {{Djayaprawira}}, \citenamefont {{Watanabe}},\
  and\ \citenamefont {{Yuasa}}}]{tulapurkar2005nt}%
  \BibitemOpen
  \bibfield  {author} {\bibinfo {author} {\bibfnamefont {A.~A.}\ \bibnamefont
  {{Tulapurkar}}}, \bibinfo {author} {\bibfnamefont {Y.}~\bibnamefont
  {{Suzuki}}}, \bibinfo {author} {\bibfnamefont {A.}~\bibnamefont
  {{Fukushima}}}, \bibinfo {author} {\bibfnamefont {H.}~\bibnamefont
  {{Kubota}}}, \bibinfo {author} {\bibfnamefont {H.}~\bibnamefont {{Maehara}}},
  \bibinfo {author} {\bibfnamefont {K.}~\bibnamefont {{Tsunekawa}}}, \bibinfo
  {author} {\bibfnamefont {D.~D.}\ \bibnamefont {{Djayaprawira}}}, \bibinfo
  {author} {\bibfnamefont {N.}~\bibnamefont {{Watanabe}}}, \ and\ \bibinfo
  {author} {\bibfnamefont {S.}~\bibnamefont {{Yuasa}}},\ }\Doi
  {10.1038/nature04207} {\bibfield  {journal} {\bibinfo  {journal} {Nature},\
  }\textbf {\bibinfo {volume} {438}},\ \bibinfo {pages} {339} (\bibinfo {year}
  {2005})}\BibitemShut {NoStop}%
\bibitem [{\citenamefont {{Deac}}\ \emph {et~al.}(2008)\citenamefont {{Deac}},
  \citenamefont {{Fukushima}}, \citenamefont {{Kubota}}, \citenamefont
  {{Maehara}}, \citenamefont {{Suzuki}}, \citenamefont {{Yuasa}}, \citenamefont
  {{Nagamine}}, \citenamefont {{Tsunekawa}}, \citenamefont {{Djayaprawira}},\
  and\ \citenamefont {{Watanabe}}}]{deac2008np}%
  \BibitemOpen
  \bibfield  {author} {\bibinfo {author} {\bibfnamefont {A.~M.}\ \bibnamefont
  {{Deac}}}, \bibinfo {author} {\bibfnamefont {A.}~\bibnamefont {{Fukushima}}},
  \bibinfo {author} {\bibfnamefont {H.}~\bibnamefont {{Kubota}}}, \bibinfo
  {author} {\bibfnamefont {H.}~\bibnamefont {{Maehara}}}, \bibinfo {author}
  {\bibfnamefont {Y.}~\bibnamefont {{Suzuki}}}, \bibinfo {author}
  {\bibfnamefont {S.}~\bibnamefont {{Yuasa}}}, \bibinfo {author} {\bibfnamefont
  {Y.}~\bibnamefont {{Nagamine}}}, \bibinfo {author} {\bibfnamefont
  {K.}~\bibnamefont {{Tsunekawa}}}, \bibinfo {author} {\bibfnamefont {D.~D.}\
  \bibnamefont {{Djayaprawira}}}, \ and\ \bibinfo {author} {\bibfnamefont
  {N.}~\bibnamefont {{Watanabe}}},\ }\Doi {10.1038/nphys1036} {\bibfield
  {journal} {\bibinfo  {journal} {Nat.\ Phys.},\ }\textbf {\bibinfo {volume}
  {4}},\ \bibinfo {pages} {803} (\bibinfo {year} {2008})}\BibitemShut {NoStop}%
\bibitem [{\citenamefont {{Nazarov}}\ \emph {et~al.}(2008)\citenamefont
  {{Nazarov}}, \citenamefont {{Nikolaev}}, \citenamefont {{Gao}}, \citenamefont
  {{Cho}},\ and\ \citenamefont {{Song}}}]{nazarov2008jap}%
  \BibitemOpen
  \bibfield  {author} {\bibinfo {author} {\bibfnamefont {A.~V.}\ \bibnamefont
  {{Nazarov}}}, \bibinfo {author} {\bibfnamefont {K.}~\bibnamefont
  {{Nikolaev}}}, \bibinfo {author} {\bibfnamefont {Z.}~\bibnamefont {{Gao}}},
  \bibinfo {author} {\bibfnamefont {H.}~\bibnamefont {{Cho}}}, \ and\ \bibinfo
  {author} {\bibfnamefont {D.}~\bibnamefont {{Song}}},\ }\Doi
  {10.1063/1.2836973} {\bibfield  {journal} {\bibinfo  {journal} {J.\ Appl.\
  Phys.},\ }\textbf {\bibinfo {volume} {103}},\ \bibinfo {pages} {070000}
  (\bibinfo {year} {2008})}\BibitemShut {NoStop}%
\bibitem [{\citenamefont {{Houssameddine}}\ \emph {et~al.}(2008)\citenamefont
  {{Houssameddine}}, \citenamefont {{Florez}}, \citenamefont {{Katine}},
  \citenamefont {{Michel}}, \citenamefont {{Ebels}}, \citenamefont {{Mauri}},
  \citenamefont {{Ozatay}}, \citenamefont {{Delaet}}, \citenamefont {{Viala}},
  \citenamefont {{Folks}}, \citenamefont {{Terris}},\ and\ \citenamefont
  {{Cyrille}}}]{houssameddine2008apl}%
  \BibitemOpen
  \bibfield  {author} {\bibinfo {author} {\bibfnamefont {D.}~\bibnamefont
  {{Houssameddine}}}, \bibinfo {author} {\bibfnamefont {S.~H.}\ \bibnamefont
  {{Florez}}}, \bibinfo {author} {\bibfnamefont {J.~A.}\ \bibnamefont
  {{Katine}}}, \bibinfo {author} {\bibfnamefont {J.-P.}\ \bibnamefont
  {{Michel}}}, \bibinfo {author} {\bibfnamefont {U.}~\bibnamefont {{Ebels}}},
  \bibinfo {author} {\bibfnamefont {D.}~\bibnamefont {{Mauri}}}, \bibinfo
  {author} {\bibfnamefont {O.}~\bibnamefont {{Ozatay}}}, \bibinfo {author}
  {\bibfnamefont {B.}~\bibnamefont {{Delaet}}}, \bibinfo {author}
  {\bibfnamefont {B.}~\bibnamefont {{Viala}}}, \bibinfo {author} {\bibfnamefont
  {L.}~\bibnamefont {{Folks}}}, \bibinfo {author} {\bibfnamefont {B.~D.}\
  \bibnamefont {{Terris}}}, \ and\ \bibinfo {author} {\bibfnamefont {M.-C.}\
  \bibnamefont {{Cyrille}}},\ }\Doi {10.1063/1.2956418} {\bibfield  {journal}
  {\bibinfo  {journal} {Appl.\ Phys.\ Lett.},\ }\textbf {\bibinfo {volume}
  {93}},\ \bibinfo {pages} {022505} (\bibinfo {year} {2008})}\BibitemShut
  {NoStop}%
\bibitem [{\citenamefont {{Zhang}}\ \emph {et~al.}(2002)\citenamefont
  {{Zhang}}, \citenamefont {{Levy}},\ and\ \citenamefont
  {{Fert}}}]{zhang2002prl}%
  \BibitemOpen
  \bibfield  {author} {\bibinfo {author} {\bibfnamefont {S.}~\bibnamefont
  {{Zhang}}}, \bibinfo {author} {\bibfnamefont {P.~M.}\ \bibnamefont {{Levy}}},
  \ and\ \bibinfo {author} {\bibfnamefont {A.}~\bibnamefont {{Fert}}},\ }\Doi
  {10.1103/PhysRevLett.88.236601} {\bibfield  {journal} {\bibinfo  {journal}
  {Phys.\ Rev.\ Lett.},\ }\textbf {\bibinfo {volume} {88}},\ \bibinfo {pages}
  {236601} (\bibinfo {year} {2002})}\BibitemShut {NoStop}%
\bibitem [{\citenamefont {{Wang}}\ \emph {et~al.}(2009)\citenamefont {{Wang}},
  \citenamefont {{Cui}}, \citenamefont {{Sun}}, \citenamefont {{Katine}},
  \citenamefont {{Buhrman}},\ and\ \citenamefont {{Ralph}}}]{wang2009prb}%
  \BibitemOpen
  \bibfield  {author} {\bibinfo {author} {\bibfnamefont {C.}~\bibnamefont
  {{Wang}}}, \bibinfo {author} {\bibfnamefont {Y.-T.}\ \bibnamefont {{Cui}}},
  \bibinfo {author} {\bibfnamefont {J.~Z.}\ \bibnamefont {{Sun}}}, \bibinfo
  {author} {\bibfnamefont {J.~A.}\ \bibnamefont {{Katine}}}, \bibinfo {author}
  {\bibfnamefont {R.~A.}\ \bibnamefont {{Buhrman}}}, \ and\ \bibinfo {author}
  {\bibfnamefont {D.~C.}\ \bibnamefont {{Ralph}}},\ }\Doi
  {10.1103/PhysRevB.79.224416} {\bibfield  {journal} {\bibinfo  {journal}
  {Phys.\ Rev.\ B},\ }\textbf {\bibinfo {volume} {79}},\ \bibinfo {pages}
  {224416} (\bibinfo {year} {2009})}\BibitemShut {NoStop}%
\bibitem [{\citenamefont {{Li}}\ \emph {et~al.}(2008)\citenamefont {{Li}},
  \citenamefont {{Zhang}}, \citenamefont {{Diao}}, \citenamefont {{Ding}},
  \citenamefont {{Tang}}, \citenamefont {{Apalkov}}, \citenamefont {{Yang}},
  \citenamefont {{Kawabata}},\ and\ \citenamefont {{Huai}}}]{li2008prl}%
  \BibitemOpen
  \bibfield  {author} {\bibinfo {author} {\bibfnamefont {Z.}~\bibnamefont
  {{Li}}}, \bibinfo {author} {\bibfnamefont {S.}~\bibnamefont {{Zhang}}},
  \bibinfo {author} {\bibfnamefont {Z.}~\bibnamefont {{Diao}}}, \bibinfo
  {author} {\bibfnamefont {Y.}~\bibnamefont {{Ding}}}, \bibinfo {author}
  {\bibfnamefont {X.}~\bibnamefont {{Tang}}}, \bibinfo {author} {\bibfnamefont
  {D.~M.}\ \bibnamefont {{Apalkov}}}, \bibinfo {author} {\bibfnamefont
  {Z.}~\bibnamefont {{Yang}}}, \bibinfo {author} {\bibfnamefont
  {K.}~\bibnamefont {{Kawabata}}}, \ and\ \bibinfo {author} {\bibfnamefont
  {Y.}~\bibnamefont {{Huai}}},\ }\Doi {10.1103/PhysRevLett.100.246602}
  {\bibfield  {journal} {\bibinfo  {journal} {Phys.\ Rev.\ Lett.},\ }\textbf
  {\bibinfo {volume} {100}},\ \bibinfo {pages} {246602} (\bibinfo {year}
  {2008})}\BibitemShut {NoStop}%
\bibitem [{\citenamefont {{Xia}}\ \emph {et~al.}(2002)\citenamefont {{Xia}},
  \citenamefont {{Kelly}}, \citenamefont {{Bauer}}, \citenamefont {{Brataas}},\
  and\ \citenamefont {{Turek}}}]{xia2002prb}%
  \BibitemOpen
  \bibfield  {author} {\bibinfo {author} {\bibfnamefont {K.}~\bibnamefont
  {{Xia}}}, \bibinfo {author} {\bibfnamefont {P.~J.}\ \bibnamefont {{Kelly}}},
  \bibinfo {author} {\bibfnamefont {G.~E.~W.}\ \bibnamefont {{Bauer}}}, \bibinfo
  {author} {\bibfnamefont {A.}~\bibnamefont {{Brataas}}}, \ and\ \bibinfo
  {author} {\bibfnamefont {I.}~\bibnamefont {{Turek}}},\ }\Doi
  {10.1103/PhysRevB.65.220401} {\bibfield  {journal} {\bibinfo  {journal}
  {Phys.\ Rev.\ B},\ }\textbf {\bibinfo {volume} {65}},\ \bibinfo {pages}
  {220401} (\bibinfo {year} {2002})}\BibitemShut {NoStop}%
\bibitem [{\citenamefont {{Zimmler}}\ \emph {et~al.}(2004)\citenamefont
  {{Zimmler}}, \citenamefont {{{\"O}zyilmaz}}, \citenamefont {{Chen}},
  \citenamefont {{Kent}}, \citenamefont {{Sun}}, \citenamefont {{Rooks}},\ and\
  \citenamefont {{Koch}}}]{zimmler2004prb}%
  \BibitemOpen
  \bibfield  {author} {\bibinfo {author} {\bibfnamefont {M.~A.}\ \bibnamefont
  {{Zimmler}}}, \bibinfo {author} {\bibfnamefont {B.}~\bibnamefont
  {{{\"O}zyilmaz}}}, \bibinfo {author} {\bibfnamefont {W.}~\bibnamefont
  {{Chen}}}, \bibinfo {author} {\bibfnamefont {A.~D.}\ \bibnamefont {{Kent}}},
  \bibinfo {author} {\bibfnamefont {J.~Z.}\ \bibnamefont {{Sun}}}, \bibinfo
  {author} {\bibfnamefont {M.~J.}\ \bibnamefont {{Rooks}}}, \ and\ \bibinfo
  {author} {\bibfnamefont {R.~H.}\ \bibnamefont {{Koch}}},\ }\Doi
  {10.1103/PhysRevB.70.184438} {\bibfield  {journal} {\bibinfo  {journal}
  {Phys.\ Rev.\ B},\ }\textbf {\bibinfo {volume} {70}},\ \bibinfo {pages}
  {184438} (\bibinfo {year} {2004})}\BibitemShut {NoStop}%
\bibitem [{\citenamefont {{Urazhdin}}\ \emph {et~al.}(2003)\citenamefont
  {{Urazhdin}}, \citenamefont {{Birge}}, \citenamefont {{Pratt}},\ and\
  \citenamefont {{Bass}}}]{urazdin2003prl}%
  \BibitemOpen
  \bibfield  {author} {\bibinfo {author} {\bibfnamefont {S.}~\bibnamefont
  {{Urazhdin}}}, \bibinfo {author} {\bibfnamefont {N.~O.}\ \bibnamefont
  {{Birge}}}, \bibinfo {author} {\bibfnamefont {W.~P.}\ \bibnamefont
  {{Pratt}}}, \ and\ \bibinfo {author} {\bibfnamefont {J.}~\bibnamefont
  {{Bass}}},\ }\Doi {10.1103/PhysRevLett.91.146803} {\bibfield  {journal}
  {\bibinfo  {journal} {Phys.\ Rev.\ Lett.},\ }\textbf {\bibinfo {volume}
  {91}},\ \bibinfo {pages} {146803} (\bibinfo {year} {2003})}\BibitemShut
  {NoStop}%
\bibitem [{\citenamefont {{Zhou}}\ and\ \citenamefont
  {{{\AA}kerman}}(2009)}]{zhou2009apl}%
  \BibitemOpen
  \bibfield  {author} {\bibinfo {author} {\bibfnamefont {Y.}~\bibnamefont
  {{Zhou}}}\ and\ \bibinfo {author} {\bibfnamefont {J.}~\bibnamefont
  {{{\AA}kerman}}},\ }\Doi {10.1063/1.3100299} {\bibfield  {journal} {\bibinfo
  {journal} {Appl.\ Phys.\ Lett.},\ }\textbf {\bibinfo {volume} {94}},\
  \bibinfo {pages} {112503} (\bibinfo {year} {2009})}\BibitemShut {NoStop}%
\bibitem [{\citenamefont {{Petit}}\ \emph {et~al.}(2007)\citenamefont
  {{Petit}}, \citenamefont {{Baraduc}}, \citenamefont {{Thirion}},
  \citenamefont {{Ebels}}, \citenamefont {{Liu}}, \citenamefont {{Li}},
  \citenamefont {{Wang}},\ and\ \citenamefont {{Dieny}}}]{petit2007prl}%
  \BibitemOpen
  \bibfield  {author} {\bibinfo {author} {\bibfnamefont {S.}~\bibnamefont
  {{Petit}}}, \bibinfo {author} {\bibfnamefont {C.}~\bibnamefont {{Baraduc}}},
  \bibinfo {author} {\bibfnamefont {C.}~\bibnamefont {{Thirion}}}, \bibinfo
  {author} {\bibfnamefont {U.}~\bibnamefont {{Ebels}}}, \bibinfo {author}
  {\bibfnamefont {Y.}~\bibnamefont {{Liu}}}, \bibinfo {author} {\bibfnamefont
  {M.}~\bibnamefont {{Li}}}, \bibinfo {author} {\bibfnamefont {P.}~\bibnamefont
  {{Wang}}}, \ and\ \bibinfo {author} {\bibfnamefont {B.}~\bibnamefont
  {{Dieny}}},\ }\Doi {10.1103/PhysRevLett.98.077203} {\bibfield  {journal}
  {\bibinfo  {journal} {Phys.\ Rev.\ Lett.},\ }\textbf {\bibinfo {volume}
  {98}},\ \bibinfo {pages} {077203} (\bibinfo {year} {2007})}\BibitemShut
  {NoStop}%
\bibitem [{\citenamefont {{Sankey}}\ \emph {et~al.}(2008)\citenamefont
  {{Sankey}}, \citenamefont {{Cui}}, \citenamefont {{Sun}}, \citenamefont
  {{Slonczewski}}, \citenamefont {{Buhrman}},\ and\ \citenamefont
  {{Ralph}}}]{sankey2008ntp}%
  \BibitemOpen
  \bibfield  {author} {\bibinfo {author} {\bibfnamefont {J.~C.}\ \bibnamefont
  {{Sankey}}}, \bibinfo {author} {\bibfnamefont {Y.-T.}\ \bibnamefont {{Cui}}},
  \bibinfo {author} {\bibfnamefont {J.~Z.}\ \bibnamefont {{Sun}}}, \bibinfo
  {author} {\bibfnamefont {J.~C.}\ \bibnamefont {{Slonczewski}}}, \bibinfo
  {author} {\bibfnamefont {R.~A.}\ \bibnamefont {{Buhrman}}}, \ and\ \bibinfo
  {author} {\bibfnamefont {D.~C.}\ \bibnamefont {{Ralph}}},\ }\Doi
  {10.1038/nphys783} {\bibfield  {journal} {\bibinfo  {journal} {Nat.\ Phys.},\
  }\textbf {\bibinfo {volume} {4}},\ \bibinfo {pages} {67} (\bibinfo {year}
  {2008})}\BibitemShut {NoStop}%
\bibitem [{\citenamefont {{Kubota}}\ \emph {et~al.}(2008)\citenamefont
  {{Kubota}}, \citenamefont {{Fukushima}}, \citenamefont {{Yakushiji}},
  \citenamefont {{Nagahama}}, \citenamefont {{Yuasa}}, \citenamefont {{Ando}},
  \citenamefont {{Maehara}}, \citenamefont {{Nagamine}}, \citenamefont
  {{Tsunekawa}}, \citenamefont {{Djayaprawira}}, \citenamefont {{Watanabe}},\
  and\ \citenamefont {{Suzuki}}}]{kubota2008ntp}%
  \BibitemOpen
  \bibfield  {author} {\bibinfo {author} {\bibfnamefont {H.}~\bibnamefont
  {{Kubota}}}, \bibinfo {author} {\bibfnamefont {A.}~\bibnamefont
  {{Fukushima}}}, \bibinfo {author} {\bibfnamefont {K.}~\bibnamefont
  {{Yakushiji}}}, \bibinfo {author} {\bibfnamefont {T.}~\bibnamefont
  {{Nagahama}}}, \bibinfo {author} {\bibfnamefont {S.}~\bibnamefont {{Yuasa}}},
  \bibinfo {author} {\bibfnamefont {K.}~\bibnamefont {{Ando}}}, \bibinfo
  {author} {\bibfnamefont {H.}~\bibnamefont {{Maehara}}}, \bibinfo {author}
  {\bibfnamefont {Y.}~\bibnamefont {{Nagamine}}}, \bibinfo {author}
  {\bibfnamefont {K.}~\bibnamefont {{Tsunekawa}}}, \bibinfo {author}
  {\bibfnamefont {D.~D.}\ \bibnamefont {{Djayaprawira}}}, \bibinfo {author}
  {\bibfnamefont {N.}~\bibnamefont {{Watanabe}}}, \ and\ \bibinfo {author}
  {\bibfnamefont {Y.}~\bibnamefont {{Suzuki}}},\ }\Doi {10.1038/nphys784}
  {\bibfield  {journal} {\bibinfo  {journal} {Nat.\ Phys.},\ }\textbf {\bibinfo
  {volume} {4}},\ \bibinfo {pages} {37} (\bibinfo {year} {2008})}\BibitemShut
  {NoStop}%
\bibitem [{\citenamefont {{Oh}}\ \emph {et~al.}(2009)\citenamefont {{Oh}},
  \citenamefont {{Park}}, \citenamefont {{Manchon}}, \citenamefont {{Chshiev}},
  \citenamefont {{Han}}, \citenamefont {{Lee}}, \citenamefont {{Lee}},
  \citenamefont {{Nam}}, \citenamefont {{Jo}}, \citenamefont {{Kong}},
  \citenamefont {{Dieny}},\ and\ \citenamefont {{Lee}}}]{oh2009ntp}%
  \BibitemOpen
  \bibfield  {author} {\bibinfo {author} {\bibfnamefont {S.}~\bibnamefont
  {{Oh}}}, \bibinfo {author} {\bibfnamefont {S.}~\bibnamefont {{Park}}},
  \bibinfo {author} {\bibfnamefont {A.}~\bibnamefont {{Manchon}}}, \bibinfo
  {author} {\bibfnamefont {M.}~\bibnamefont {{Chshiev}}}, \bibinfo {author}
  {\bibfnamefont {J.}~\bibnamefont {{Han}}}, \bibinfo {author} {\bibfnamefont
  {H.}~\bibnamefont {{Lee}}}, \bibinfo {author} {\bibfnamefont
  {J.}~\bibnamefont {{Lee}}}, \bibinfo {author} {\bibfnamefont
  {K.}~\bibnamefont {{Nam}}}, \bibinfo {author} {\bibfnamefont
  {Y.}~\bibnamefont {{Jo}}}, \bibinfo {author} {\bibfnamefont {Y.}~\bibnamefont
  {{Kong}}}, \bibinfo {author} {\bibfnamefont {B.}~\bibnamefont {{Dieny}}}, \
  and\ \bibinfo {author} {\bibfnamefont {K.}~\bibnamefont {{Lee}}},\ }\Doi
  {10.1038/nphys1427} {\bibfield  {journal} {\bibinfo  {journal} {Nat.\
  Phys.},\ }\textbf {\bibinfo {volume} {5}},\ \bibinfo {pages} {898} (\bibinfo
  {year} {2009})}\BibitemShut {NoStop}%
\bibitem [{\citenamefont {{Heinonen}}(2010)}]{heinonen2010prb}%
  \BibitemOpen
  \bibfield  {author} {\bibinfo {author} {\bibfnamefont {O.~G.}\ \bibnamefont
  {{Heinonen}}},\ }\Doi {10.1103/PhysRevB.81.054405} {\bibfield  {journal}
  {\bibinfo  {journal} {Phys.\ Rev.\ B},\ }\textbf {\bibinfo {volume} {81}},\
  \bibinfo {pages} {054405} (\bibinfo {year} {2010})}\BibitemShut {NoStop}%
\bibitem [{\citenamefont {{Heinonen}}\ \emph {et~al.}(2010)\citenamefont
  {{Heinonen}}, \citenamefont {{Stokes}},\ and\ \citenamefont
  {{Yi}}}]{heinonen2010prl}%
  \BibitemOpen
  \bibfield  {author} {\bibinfo {author} {\bibfnamefont {O.~G.}\ \bibnamefont
  {{Heinonen}}}, \bibinfo {author} {\bibfnamefont {S.~W.}\ \bibnamefont
  {{Stokes}}}, \ and\ \bibinfo {author} {\bibfnamefont {J.~Y.}\ \bibnamefont
  {{Yi}}},\ }\Doi {10.1103/PhysRevLett.105.066602} {\bibfield  {journal}
  {\bibinfo  {journal} {Phys.\ Rev.\ Lett.},\ }\textbf {\bibinfo {volume}
  {105}},\ \bibinfo {pages} {066602} (\bibinfo {year} {2010})}\BibitemShut
  {NoStop}%
\bibitem [{\citenamefont {{Jung}}\ \emph {et~al.}(2010)\citenamefont {{Jung}},
  \citenamefont {{Park}}, \citenamefont {{You}},\ and\ \citenamefont
  {{Yuasa}}}]{jung2010prb}%
  \BibitemOpen
  \bibfield  {author} {\bibinfo {author} {\bibfnamefont {M.~H.}\ \bibnamefont
  {{Jung}}}, \bibinfo {author} {\bibfnamefont {S.}~\bibnamefont {{Park}}},
  \bibinfo {author} {\bibfnamefont {C.~Y.}~\bibnamefont {{You}}}, \ and\ \bibinfo
  {author} {\bibfnamefont {S.}~\bibnamefont {{Yuasa}}},\ }\Doi
  {10.1103/PhysRevB.81.134419} {\bibfield  {journal} {\bibinfo  {journal}
  {Phys.\ Rev.\ B},\ }\textbf {\bibinfo {volume} {81}},\ \bibinfo {pages}
  {134419} (\bibinfo {year} {2010})}\BibitemShut {NoStop}%
\bibitem [{\citenamefont {{Theodonis}}\ \emph {et~al.}(2006)\citenamefont
  {{Theodonis}}, \citenamefont {{Kioussis}}, \citenamefont {{Kalitsov}},
  \citenamefont {{Chshiev}},\ and\ \citenamefont
  {{Butler}}}]{theodonis2006prl}%
  \BibitemOpen
  \bibfield  {author} {\bibinfo {author} {\bibfnamefont {I.}~\bibnamefont
  {{Theodonis}}}, \bibinfo {author} {\bibfnamefont {N.}~\bibnamefont
  {{Kioussis}}}, \bibinfo {author} {\bibfnamefont {A.}~\bibnamefont
  {{Kalitsov}}}, \bibinfo {author} {\bibfnamefont {M.}~\bibnamefont
  {{Chshiev}}}, \ and\ \bibinfo {author} {\bibfnamefont {W.~H.}\ \bibnamefont
  {{Butler}}},\ }\Doi {10.1103/PhysRevLett.97.237205} {\bibfield  {journal}
  {\bibinfo  {journal} {Phys.\ Rev.\ Lett.},\ }\textbf {\bibinfo {volume}
  {97}},\ \bibinfo {pages} {237205} (\bibinfo {year} {2006})}\BibitemShut
  {NoStop}%
\bibitem [{\citenamefont {{Heiliger}}\ and\ \citenamefont
  {{Stiles}}(2008)}]{stiles2008prb}%
  \BibitemOpen
  \bibfield  {author} {\bibinfo {author} {\bibfnamefont {C.}~\bibnamefont
  {{Heiliger}}}\ and\ \bibinfo {author} {\bibfnamefont {M.~D.}\ \bibnamefont
  {{Stiles}}},\ }\Doi {10.1103/PhysRevLett.100.186805} {\bibfield  {journal}
  {\bibinfo  {journal} {Phys.\ Rev.\ Lett.},\ }\textbf {\bibinfo {volume}
  {100}},\ \bibinfo {pages} {186805} (\bibinfo {year} {2008})}\BibitemShut
  {NoStop}%
\bibitem [{\citenamefont {{Xiao}}\ \emph {et~al.}(2008)\citenamefont {{Xiao}},
  \citenamefont {{Bauer}},\ and\ \citenamefont {{Brataas}}}]{xiao2008prb}%
  \BibitemOpen
  \bibfield  {author} {\bibinfo {author} {\bibfnamefont {J.}~\bibnamefont
  {{Xiao}}}, \bibinfo {author} {\bibfnamefont {G.~E.~W.}\ \bibnamefont
  {{Bauer}}}, \ and\ \bibinfo {author} {\bibfnamefont {A.}~\bibnamefont
  {{Brataas}}},\ }\Doi {10.1103/PhysRevB.77.224419} {\bibfield  {journal}
  {\bibinfo  {journal} {Phys.\ Rev.\ B},\ }\textbf {\bibinfo {volume} {77}},\
  \bibinfo {pages} {224419} (\bibinfo {year} {2008})}\BibitemShut {NoStop}%
\bibitem [{\citenamefont {{Cornelissen}}\ \emph {et~al.}(2009)\citenamefont
  {{Cornelissen}}, \citenamefont {{Bianchini}}, \citenamefont {{Hrkac}},
  \citenamefont {{de Beeck}}, \citenamefont {{Lagae}}, \citenamefont {{Kim}},
  \citenamefont {{Devolder}}, \citenamefont {{Crozat}}, \citenamefont
  {{Chappert}},\ and\ \citenamefont {{Schrefl}}}]{cornelissen2009epl}%
  \BibitemOpen
  \bibfield  {author} {\bibinfo {author} {\bibfnamefont {S.}~\bibnamefont
  {{Cornelissen}}}, \bibinfo {author} {\bibfnamefont {L.}~\bibnamefont
  {{Bianchini}}}, \bibinfo {author} {\bibfnamefont {G.}~\bibnamefont
  {{Hrkac}}}, \bibinfo {author} {\bibfnamefont {M.~O.}\ \bibnamefont {{de
  Beeck}}}, \bibinfo {author} {\bibfnamefont {L.}~\bibnamefont {{Lagae}}},
  \bibinfo {author} {\bibfnamefont {J.~V.}~\bibnamefont {{Kim}}}, \bibinfo
  {author} {\bibfnamefont {T.}~\bibnamefont {{Devolder}}}, \bibinfo {author}
  {\bibfnamefont {P.}~\bibnamefont {{Crozat}}}, \bibinfo {author}
  {\bibfnamefont {C.}~\bibnamefont {{Chappert}}}, \ and\ \bibinfo {author}
  {\bibfnamefont {T.}~\bibnamefont {{Schrefl}}},\ }\Doi
  {10.1209/0295-5075/87/57001} {\bibfield  {journal} {\bibinfo  {journal}
  {Europhys. Lett.},\ }\textbf {\bibinfo {volume} {87}},\ \bibinfo {pages}
  {57001} (\bibinfo {year} {2009})}\BibitemShut {NoStop}%
\bibitem [{\citenamefont {{Cornelissen}}\ \emph {et~al.}(2010)\citenamefont
  {{Cornelissen}}, \citenamefont {{Bianchini}}, \citenamefont {{Devolder}},
  \citenamefont {{Kim}}, \citenamefont {{van Roy}}, \citenamefont {{Lagae}},\
  and\ \citenamefont {{Chappert}}}]{cornelissen2010prb}%
  \BibitemOpen
  \bibfield  {author} {\bibinfo {author} {\bibfnamefont {S.}~\bibnamefont
  {{Cornelissen}}}, \bibinfo {author} {\bibfnamefont {L.}~\bibnamefont
  {{Bianchini}}}, \bibinfo {author} {\bibfnamefont {T.}~\bibnamefont
  {{Devolder}}}, \bibinfo {author} {\bibfnamefont {J.~V.}~\bibnamefont {{Kim}}},
  \bibinfo {author} {\bibfnamefont {W.}~\bibnamefont {{VanRoy}}}, \bibinfo
  {author} {\bibfnamefont {L.}~\bibnamefont {{Lagae}}}, \ and\ \bibinfo
  {author} {\bibfnamefont {C.}~\bibnamefont {{Chappert}}},\ }\Doi
  {10.1103/PhysRevB.81.144408} {\bibfield  {journal} {\bibinfo  {journal}
  {Phys.\ Rev.\ B},\ }\textbf {\bibinfo {volume} {81}},\ \bibinfo {pages}
  {144408} (\bibinfo {year} {2010})}\BibitemShut {NoStop}%
\bibitem [{\citenamefont {{Houssameddine}}\ \emph {et~al.}(2010)\citenamefont
  {{Houssameddine}}, \citenamefont {{Sierra}}, \citenamefont {{Gusakova}},
  \citenamefont {{Delaet}}, \citenamefont {{Ebels}}, \citenamefont
  {{Buda-Prejbeanu}}, \citenamefont {{Cyrille}}, \citenamefont {{Dieny}},
  \citenamefont {{Ocker}}, \citenamefont {{Langer}},\ and\ \citenamefont
  {{Maas}}}]{houssameddine2010apl}%
  \BibitemOpen
  \bibfield  {author} {\bibinfo {author} {\bibfnamefont {D.}~\bibnamefont
  {{Houssameddine}}}, \bibinfo {author} {\bibfnamefont {J.~F.}\ \bibnamefont
  {{Sierra}}}, \bibinfo {author} {\bibfnamefont {D.}~\bibnamefont
  {{Gusakova}}}, \bibinfo {author} {\bibfnamefont {B.}~\bibnamefont
  {{Delaet}}}, \bibinfo {author} {\bibfnamefont {U.}~\bibnamefont {{Ebels}}},
  \bibinfo {author} {\bibfnamefont {L.~D.}\ \bibnamefont {{Buda-Prejbeanu}}},
  \bibinfo {author} {\bibfnamefont {M.}~\bibnamefont {{Cyrille}}}, \bibinfo
  {author} {\bibfnamefont {B.}~\bibnamefont {{Dieny}}}, \bibinfo {author}
  {\bibfnamefont {B.}~\bibnamefont {{Ocker}}}, \bibinfo {author} {\bibfnamefont
  {J.}~\bibnamefont {{Langer}}}, \ and\ \bibinfo {author} {\bibfnamefont
  {W.}~\bibnamefont {{Maas}}},\ }\Doi {10.1063/1.3314282} {\bibfield  {journal}
  {\bibinfo  {journal} {Appl.\ Phys.\ Lett.},\ }\textbf {\bibinfo {volume}
  {96}},\ \bibinfo {pages} {072511} (\bibinfo {year} {2010})}\BibitemShut
  {NoStop}%
\bibitem [{\citenamefont {{Mao}}\ \emph {et~al.}(2006)\citenamefont {{Mao}},
  \citenamefont {{Chen}}, \citenamefont {{Liu}}, \citenamefont {{Chen}},
  \citenamefont {{Xu}}, \citenamefont {{Lu}}, \citenamefont {{Patwari}},
  \citenamefont {{Xi}}, \citenamefont {{Chang}}, \citenamefont {{Miller}},
  \citenamefont {{Menard}}, \citenamefont {{Pant}}, \citenamefont {{Loven}},
  \citenamefont {{Duxstad}}, \citenamefont {{Li}}, \citenamefont {{Zhang}},
  \citenamefont {{Johnston}}, \citenamefont {{Lamberton}}, \citenamefont
  {{Gubbins}}, \citenamefont {{McLaughlin}}, \citenamefont {{Gadbois}},
  \citenamefont {{Ding}}, \citenamefont {{Cross}}, \citenamefont {{Xue}},\ and\
  \citenamefont {{Ryan}}}]{mao2006ieeem}%
  \BibitemOpen
  \bibfield  {author} {\bibinfo {author} {\bibfnamefont {S.}~\bibnamefont
  {{Mao}}}, \bibinfo {author} {\bibfnamefont {Y.}~\bibnamefont {{Chen}}},
  \bibinfo {author} {\bibfnamefont {F.}~\bibnamefont {{Liu}}}, \bibinfo
  {author} {\bibfnamefont {X.}~\bibnamefont {{Chen}}}, \bibinfo {author}
  {\bibfnamefont {B.}~\bibnamefont {{Xu}}}, \bibinfo {author} {\bibfnamefont
  {C.~P.}\ \bibnamefont {{Lu}}}, \bibinfo {author} {\bibfnamefont
  {M.}~\bibnamefont {{Patwari}}}, \bibinfo {author} {\bibfnamefont
  {H.}~\bibnamefont {{Xi}}}, \bibinfo {author} {\bibfnamefont {C.}~\bibnamefont
  {{Chang}}}, \bibinfo {author} {\bibfnamefont {B.}~\bibnamefont {{Miller}}},
  \bibinfo {author} {\bibfnamefont {D.}~\bibnamefont {{Menard}}}, \bibinfo
  {author} {\bibfnamefont {B.}~\bibnamefont {{Pant}}}, \bibinfo {author}
  {\bibfnamefont {J.}~\bibnamefont {{Loven}}}, \bibinfo {author} {\bibfnamefont
  {K.}~\bibnamefont {{Duxstad}}}, \bibinfo {author} {\bibfnamefont
  {S.}~\bibnamefont {{Li}}}, \bibinfo {author} {\bibfnamefont {Z.}~\bibnamefont
  {{Zhang}}}, \bibinfo {author} {\bibfnamefont {A.}~\bibnamefont {{Johnston}}},
  \bibinfo {author} {\bibfnamefont {R.}~\bibnamefont {{Lamberton}}}, \bibinfo
  {author} {\bibfnamefont {M.}~\bibnamefont {{Gubbins}}}, \bibinfo {author}
  {\bibfnamefont {T.}~\bibnamefont {{McLaughlin}}}, \bibinfo {author}
  {\bibfnamefont {J.}~\bibnamefont {{Gadbois}}}, \bibinfo {author}
  {\bibfnamefont {J.}~\bibnamefont {{Ding}}}, \bibinfo {author} {\bibfnamefont
  {B.}~\bibnamefont {{Cross}}}, \bibinfo {author} {\bibfnamefont
  {S.}~\bibnamefont {{Xue}}}, \ and\ \bibinfo {author} {\bibfnamefont
  {P.}~\bibnamefont {{Ryan}}},\ }\Doi {10.1109/TMAG.2005.861788} {\bibfield
  {journal} {\bibinfo  {journal} {IEEE Trans. Magn.},\ }\textbf {\bibinfo
  {volume} {42}},\ \bibinfo {pages} {97} (\bibinfo {year} {2006})}\BibitemShut
  {NoStop}%
\bibitem [{\citenamefont {{Nazarov}}\ \emph {et~al.}(2006)\citenamefont
  {{Nazarov}}, \citenamefont {{Olson}}, \citenamefont {{Cho}}, \citenamefont
  {{Nikolaev}}, \citenamefont {{Gao}}, \citenamefont {{Stokes}},\ and\
  \citenamefont {{Pant}}}]{nazarov2006apl}%
  \BibitemOpen
  \bibfield  {author} {\bibinfo {author} {\bibfnamefont {A.~V.}\ \bibnamefont
  {{Nazarov}}}, \bibinfo {author} {\bibfnamefont {H.~M.}\ \bibnamefont
  {{Olson}}}, \bibinfo {author} {\bibfnamefont {H.}~\bibnamefont {{Cho}}},
  \bibinfo {author} {\bibfnamefont {K.}~\bibnamefont {{Nikolaev}}}, \bibinfo
  {author} {\bibfnamefont {Z.}~\bibnamefont {{Gao}}}, \bibinfo {author}
  {\bibfnamefont {S.}~\bibnamefont {{Stokes}}}, \ and\ \bibinfo {author}
  {\bibfnamefont {B.~B.}\ \bibnamefont {{Pant}}},\ }\Doi {10.1063/1.2196232}
  {\bibfield  {journal} {\bibinfo  {journal} {Appl.\ Phys.\ Lett.},\ }\textbf
  {\bibinfo {volume} {88}},\ \bibinfo {pages} {162504} (\bibinfo {year}
  {2006})}\BibitemShut {NoStop}%
\bibitem [{\citenamefont {{Kubota}}\ \emph {et~al.}(2006)\citenamefont
  {{Kubota}}, \citenamefont {{Fukushima}}, \citenamefont {{Ootani}},
  \citenamefont {{Yuasa}}, \citenamefont {{Ando}}, \citenamefont {{Maehara}},
  \citenamefont {{Tsunekawa}}, \citenamefont {{Djayaprawira}}, \citenamefont
  {{Watanabe}},\ and\ \citenamefont {{Suzuki}}}]{kubota2006apl}%
  \BibitemOpen
  \bibfield  {author} {\bibinfo {author} {\bibfnamefont {H.}~\bibnamefont
  {{Kubota}}}, \bibinfo {author} {\bibfnamefont {A.}~\bibnamefont
  {{Fukushima}}}, \bibinfo {author} {\bibfnamefont {Y.}~\bibnamefont
  {{Ootani}}}, \bibinfo {author} {\bibfnamefont {S.}~\bibnamefont {{Yuasa}}},
  \bibinfo {author} {\bibfnamefont {K.}~\bibnamefont {{Ando}}}, \bibinfo
  {author} {\bibfnamefont {H.}~\bibnamefont {{Maehara}}}, \bibinfo {author}
  {\bibfnamefont {K.}~\bibnamefont {{Tsunekawa}}}, \bibinfo {author}
  {\bibfnamefont {D.~D.}\ \bibnamefont {{Djayaprawira}}}, \bibinfo {author}
  {\bibfnamefont {N.}~\bibnamefont {{Watanabe}}}, \ and\ \bibinfo {author}
  {\bibfnamefont {Y.}~\bibnamefont {{Suzuki}}},\ }\Doi {10.1063/1.2222241}
  {\bibfield  {journal} {\bibinfo  {journal} {Appl.\ Phys.\ Lett.},\ }\textbf
  {\bibinfo {volume} {89}},\ \bibinfo {pages} {032505} (\bibinfo {year}
  {2006})}\BibitemShut {NoStop}%
\bibitem [{\citenamefont {{Huai}}\ \emph {et~al.}(2006)\citenamefont {{Huai}},
  \citenamefont {{Apalkov}}, \citenamefont {{Diao}}, \citenamefont {{Ding}},
  \citenamefont {{Panchula}}, \citenamefont {{Pakala}}, \citenamefont
  {{Wang}},\ and\ \citenamefont {{Chen}}}]{huai2006jjap}%
  \BibitemOpen
  \bibfield  {author} {\bibinfo {author} {\bibfnamefont {Y.}~\bibnamefont
  {{Huai}}}, \bibinfo {author} {\bibfnamefont {D.}~\bibnamefont {{Apalkov}}},
  \bibinfo {author} {\bibfnamefont {Z.}~\bibnamefont {{Diao}}}, \bibinfo
  {author} {\bibfnamefont {Y.}~\bibnamefont {{Ding}}}, \bibinfo {author}
  {\bibfnamefont {A.}~\bibnamefont {{Panchula}}}, \bibinfo {author}
  {\bibfnamefont {M.}~\bibnamefont {{Pakala}}}, \bibinfo {author}
  {\bibfnamefont {L.}~\bibnamefont {{Wang}}}, \ and\ \bibinfo {author}
  {\bibfnamefont {E.}~\bibnamefont {{Chen}}},\ }\Doi {10.1143/JJAP.45.3835}
  {\bibfield  {journal} {\bibinfo  {journal} {Jpn.\ J.\ Appl.\ Phys.},\
  }\textbf {\bibinfo {volume} {45}},\ \bibinfo {pages} {3835} (\bibinfo {year}
  {2006})}\BibitemShut {NoStop}%
\bibitem [{\citenamefont {{Bonetti}}\ \emph {et~al.}(2009)\citenamefont
  {{Bonetti}}, \citenamefont {{Muduli}}, \citenamefont {{Mancoff}},\ and\
  \citenamefont {{{\AA}kerman}}}]{bonetti2009apl}%
  \BibitemOpen
  \bibfield  {author} {\bibinfo {author} {\bibfnamefont {S.}~\bibnamefont
  {{Bonetti}}}, \bibinfo {author} {\bibfnamefont {P.}~\bibnamefont {{Muduli}}},
  \bibinfo {author} {\bibfnamefont {F.}~\bibnamefont {{Mancoff}}}, \ and\
  \bibinfo {author} {\bibfnamefont {J.}~\bibnamefont {{{\AA}kerman}}},\ }\Doi
  {10.1063/1.3097238} {\bibfield  {journal} {\bibinfo  {journal} {Appl.\ Phys.\
  Lett.},\ }\textbf {\bibinfo {volume} {94}},\ \bibinfo {pages} {102507}
  (\bibinfo {year} {2009})}\BibitemShut {NoStop}%
\bibitem [{\citenamefont {{J{\"o}nsson-{\AA}Kerman}}\ \emph
  {et~al.}(2000)\citenamefont {{J{\"o}nsson-{\AA}Kerman}}, \citenamefont
  {{Escudero}}, \citenamefont {{Leighton}}, \citenamefont {{Kim}},
  \citenamefont {{Schuller}},\ and\ \citenamefont {{Rabson}}}]{akerman2000apl}%
  \BibitemOpen
  \bibfield  {author} {\bibinfo {author} {\bibfnamefont {B.~J.}\ \bibnamefont
  {{J{\"o}nsson-{\AA}Kerman}}}, \bibinfo {author} {\bibfnamefont
  {R.}~\bibnamefont {{Escudero}}}, \bibinfo {author} {\bibfnamefont
  {C.}~\bibnamefont {{Leighton}}}, \bibinfo {author} {\bibfnamefont
  {S.}~\bibnamefont {{Kim}}}, \bibinfo {author} {\bibfnamefont {I.~K.}\
  \bibnamefont {{Schuller}}}, \ and\ \bibinfo {author} {\bibfnamefont {D.~A.}\
  \bibnamefont {{Rabson}}},\ }\Doi {10.1063/1.1310633} {\bibfield  {journal}
  {\bibinfo  {journal} {Appl.\ Phys.\ Lett.},\ }\textbf {\bibinfo {volume}
  {77}},\ \bibinfo {pages} {1870} (\bibinfo {year} {2000})}\BibitemShut
  {NoStop}%
\bibitem [{\citenamefont {{Rabson}}\ \emph {et~al.}(2001)\citenamefont
  {{Rabson}}, \citenamefont {{J{\"o}nsson-{\AA}Kerman}}, \citenamefont
  {{Romero}}, \citenamefont {{Escudero}}, \citenamefont {{Leighton}},
  \citenamefont {{Kim}},\ and\ \citenamefont {{Schuller}}}]{rabson2001jap}%
  \BibitemOpen
  \bibfield  {author} {\bibinfo {author} {\bibfnamefont {D.~A.}\ \bibnamefont
  {{Rabson}}}, \bibinfo {author} {\bibfnamefont {B.~J.}\ \bibnamefont
  {{J{\"o}nsson-{\AA}Kerman}}}, \bibinfo {author} {\bibfnamefont {A.~H.}\
  \bibnamefont {{Romero}}}, \bibinfo {author} {\bibfnamefont {R.}~\bibnamefont
  {{Escudero}}}, \bibinfo {author} {\bibfnamefont {C.}~\bibnamefont
  {{Leighton}}}, \bibinfo {author} {\bibfnamefont {S.}~\bibnamefont {{Kim}}}, \
  and\ \bibinfo {author} {\bibfnamefont {I.~K.}\ \bibnamefont {{Schuller}}},\
  }\Doi {10.1063/1.1344220} {\bibfield  {journal} {\bibinfo  {journal} {J.\
  Appl.\ Phys.},\ }\textbf {\bibinfo {volume} {89}},\ \bibinfo {pages} {2786}
  (\bibinfo {year} {2001})}\BibitemShut {NoStop}%
\bibitem [{\citenamefont {{{\AA}kerman}}\ \emph {et~al.}(2001)\citenamefont
  {{{\AA}kerman}}, \citenamefont {{Slaughter}}, \citenamefont {{Dave}},\ and\
  \citenamefont {{Schuller}}}]{akerman2001apl}%
  \BibitemOpen
  \bibfield  {author} {\bibinfo {author} {\bibfnamefont {J.~J.}\ \bibnamefont
  {{{\AA}kerman}}}, \bibinfo {author} {\bibfnamefont {J.~M.}\ \bibnamefont
  {{Slaughter}}}, \bibinfo {author} {\bibfnamefont {R.~W.}\ \bibnamefont
  {{Dave}}}, \ and\ \bibinfo {author} {\bibfnamefont {I.~K.}\ \bibnamefont
  {{Schuller}}},\ }\Doi {10.1063/1.1413716} {\bibfield  {journal} {\bibinfo
  {journal} {Appl.\ Phys.\ Lett.},\ }\textbf {\bibinfo {volume} {79}},\
  \bibinfo {pages} {3104} (\bibinfo {year} {2001})}\BibitemShut {NoStop}%
\bibitem [{\citenamefont {{{\AA}kerman}}\ \emph {et~al.}(2002)\citenamefont
  {{{\AA}kerman}}, \citenamefont {{Escudero}}, \citenamefont {{Leighton}},
  \citenamefont {{Kim}}, \citenamefont {{Rabson}}, \citenamefont {{Dave}},
  \citenamefont {{Slaughter}},\ and\ \citenamefont
  {{Schuller}}}]{akerman2002jmmm}%
  \BibitemOpen
  \bibfield  {author} {\bibinfo {author} {\bibfnamefont {J.~J.}\ \bibnamefont
  {{{\AA}kerman}}}, \bibinfo {author} {\bibfnamefont {R.}~\bibnamefont
  {{Escudero}}}, \bibinfo {author} {\bibfnamefont {C.}~\bibnamefont
  {{Leighton}}}, \bibinfo {author} {\bibfnamefont {S.}~\bibnamefont {{Kim}}},
  \bibinfo {author} {\bibfnamefont {D.~A.}\ \bibnamefont {{Rabson}}}, \bibinfo
  {author} {\bibfnamefont {R.~W.}\ \bibnamefont {{Dave}}}, \bibinfo {author}
  {\bibfnamefont {J.~M.}\ \bibnamefont {{Slaughter}}}, \ and\ \bibinfo {author}
  {\bibfnamefont {I.~K.}\ \bibnamefont {{Schuller}}},\ }\Doi
  {10.1016/S0304-8853(01)00712-0} {\bibfield  {journal} {\bibinfo  {journal}
  {J.\ Magn.\ Magn.\ Mater.},\ }\textbf {\bibinfo {volume} {240}},\ \bibinfo
  {pages} {86} (\bibinfo {year} {2002})}\BibitemShut {NoStop}%
\bibitem [{\citenamefont {{Teixeira}}\ \emph {et~al.}(2009)\citenamefont
  {{Teixeira}}, \citenamefont {{Ventura}}, \citenamefont {{Carpinteiro}},
  \citenamefont {{Araujo}}, \citenamefont {{Sousa}}, \citenamefont
  {{Wisniowski}},\ and\ \citenamefont {{Freitas}}}]{teixeira2009jap}%
  \BibitemOpen
  \bibfield  {author} {\bibinfo {author} {\bibfnamefont {J.~M.}\ \bibnamefont
  {{Teixeira}}}, \bibinfo {author} {\bibfnamefont {J.}~\bibnamefont
  {{Ventura}}}, \bibinfo {author} {\bibfnamefont {F.}~\bibnamefont
  {{Carpinteiro}}}, \bibinfo {author} {\bibfnamefont {J.~P.}\ \bibnamefont
  {{Araujo}}}, \bibinfo {author} {\bibfnamefont {J.~B.}\ \bibnamefont
  {{Sousa}}}, \bibinfo {author} {\bibfnamefont {P.}~\bibnamefont
  {{Wisniowski}}}, \ and\ \bibinfo {author} {\bibfnamefont {P.~P.}\
  \bibnamefont {{Freitas}}},\ }\Doi {10.1063/1.3236512} {\bibfield  {journal}
  {\bibinfo  {journal} {J.\ Appl.\ Phys.},\ }\textbf {\bibinfo {volume}
  {106}},\ \bibinfo {pages} {073707} (\bibinfo {year} {2009})}\BibitemShut
  {NoStop}%
\bibitem [{\citenamefont {{Fuchs}}\ \emph {et~al.}(2006)\citenamefont
  {{Fuchs}}, \citenamefont {{Katine}}, \citenamefont {{Kiselev}}, \citenamefont
  {{Mauri}}, \citenamefont {{Wooley}}, \citenamefont {{Ralph}},\ and\
  \citenamefont {{Buhrman}}}]{fuchs2006prl}%
  \BibitemOpen
  \bibfield  {author} {\bibinfo {author} {\bibfnamefont {G.~D.}\ \bibnamefont
  {{Fuchs}}}, \bibinfo {author} {\bibfnamefont {J.~A.}\ \bibnamefont
  {{Katine}}}, \bibinfo {author} {\bibfnamefont {S.~I.}\ \bibnamefont
  {{Kiselev}}}, \bibinfo {author} {\bibfnamefont {D.}~\bibnamefont {{Mauri}}},
  \bibinfo {author} {\bibfnamefont {K.~S.}\ \bibnamefont {{Wooley}}}, \bibinfo
  {author} {\bibfnamefont {D.~C.}\ \bibnamefont {{Ralph}}}, \ and\ \bibinfo
  {author} {\bibfnamefont {R.~A.}\ \bibnamefont {{Buhrman}}},\ }\Doi
  {10.1103/PhysRevLett.96.186603} {\bibfield  {journal} {\bibinfo  {journal}
  {Phys.\ Rev.\ Lett.},\ }\textbf {\bibinfo {volume} {96}},\ \bibinfo {pages}
  {186603} (\bibinfo {year} {2006})}\BibitemShut {NoStop}%
\bibitem [{\citenamefont {{{\AA}kerman}}\ \emph {et~al.}(2001)\citenamefont
  {{{\AA}kerman}}, \citenamefont {{Roshchin}},\citenamefont {{Slaughter}}, \citenamefont {{Dave}},\ and\
  \citenamefont {{Schuller}}}]{akerman2003epl}%
  \BibitemOpen
  \bibfield  {author} {\bibinfo {author} {\bibfnamefont {J.~J.}\ \bibnamefont
  {{{\AA}kerman}}}, \bibinfo {author} {\bibfnamefont {I.~V.}\ \bibnamefont
  {{Roshchin}}}, \bibinfo {author} {\bibfnamefont {J.~M.}\ \bibnamefont
  {{Slaughter}}}, \ \bibinfo {author} {\bibfnamefont {R.~W.}\ \bibnamefont
  {{Dave}}}, \ and\ \bibinfo {author} {\bibfnamefont {I.~K.}\ \bibnamefont
  {{Schuller}}},\ }\Doi {10.1209/epl/i2003-00484-4} {\bibfield  {journal} {\bibinfo
  {journal} {Europhys.\ Lett.},\ }\textbf {\bibinfo {volume} {63}},\
  \bibinfo {pages} {104} (\bibinfo {year} {2003})}\BibitemShut {NoStop}%
\bibitem [{\citenamefont {{Parkin}}\ \emph {et~al.}(1999)\citenamefont
  {{Parkin}}, \citenamefont {{Roche}}, \citenamefont {{Samant}}, \citenamefont
  {{Rice}}, \citenamefont {{Beyers}}, \citenamefont {{Scheuerlein}},
  \citenamefont {{O'Sullivan}}, \citenamefont {{Brown}}, \citenamefont
  {{Bucchigano}}, \citenamefont {{Abraham}}, \citenamefont {{Lu}},
  \citenamefont {{Rooks}}, \citenamefont {{Trouilloud}}, \citenamefont
  {{Wanner}},\ and\ \citenamefont {{Gallagher}}}]{parkin1999jap}%
  \BibitemOpen
  \bibfield  {author} {\bibinfo {author} {\bibfnamefont {S.~S.~P.}\
  \bibnamefont {{Parkin}}}, \bibinfo {author} {\bibfnamefont {K.~P.}\
  \bibnamefont {{Roche}}}, \bibinfo {author} {\bibfnamefont {M.~G.}\
  \bibnamefont {{Samant}}}, \bibinfo {author} {\bibfnamefont {P.~M.}\
  \bibnamefont {{Rice}}}, \bibinfo {author} {\bibfnamefont {R.~B.}\
  \bibnamefont {{Beyers}}}, \bibinfo {author} {\bibfnamefont {R.~E.}\
  \bibnamefont {{Scheuerlein}}}, \bibinfo {author} {\bibfnamefont {E.~J.}\
  \bibnamefont {{O'Sullivan}}}, \bibinfo {author} {\bibfnamefont {S.~L.}\
  \bibnamefont {{Brown}}}, \bibinfo {author} {\bibfnamefont {J.}~\bibnamefont
  {{Bucchigano}}}, \bibinfo {author} {\bibfnamefont {D.~W.}\ \bibnamefont
  {{Abraham}}}, \bibinfo {author} {\bibfnamefont {Y.}~\bibnamefont {{Lu}}},
  \bibinfo {author} {\bibfnamefont {M.}~\bibnamefont {{Rooks}}}, \bibinfo
  {author} {\bibfnamefont {P.~L.}\ \bibnamefont {{Trouilloud}}}, \bibinfo
  {author} {\bibfnamefont {R.~A.}\ \bibnamefont {{Wanner}}}, \ and\ \bibinfo
  {author} {\bibfnamefont {W.~J.}\ \bibnamefont {{Gallagher}}},\ }\Doi
  {10.1063/1.369932} {\bibfield  {journal} {\bibinfo  {journal} {J.\ Appl.\
  Phys.},\ }\textbf {\bibinfo {volume} {85}},\ \bibinfo {pages} {5828}
  (\bibinfo {year} {1999})}\BibitemShut {NoStop}%
\bibitem [{\citenamefont {{Helmer}}\ \emph {et~al.}(2010)\citenamefont
  {{Helmer}}, \citenamefont {{Cornelissen}}, \citenamefont {{Devolder}},
  \citenamefont {{Kim}}, \citenamefont {{van Roy}}, \citenamefont {{Lagae}},\
  and\ \citenamefont {{Chappert}}}]{helmer2010prb}%
  \BibitemOpen
  \bibfield  {author} {\bibinfo {author} {\bibfnamefont {A.}~\bibnamefont
  {{Helmer}}}, \bibinfo {author} {\bibfnamefont {S.}~\bibnamefont
  {{Cornelissen}}}, \bibinfo {author} {\bibfnamefont {T.}~\bibnamefont
  {{Devolder}}}, \bibinfo {author} {\bibfnamefont {J.~V.}~\bibnamefont {{Kim}}},
  \bibinfo {author} {\bibfnamefont {W.}~\bibnamefont {{vanRoy}}}, \bibinfo
  {author} {\bibfnamefont {L.}~\bibnamefont {{Lagae}}}, \ and\ \bibinfo
  {author} {\bibfnamefont {C.}~\bibnamefont {{Chappert}}},\ }\Doi
  {10.1103/PhysRevB.81.094416} {\bibfield  {journal} {\bibinfo  {journal}
  {Phys.\ Rev.\ B},\ }\textbf {\bibinfo {volume} {81}},\ \bibinfo {pages}
  {094416} (\bibinfo {year} {2010})}\BibitemShut {NoStop}%
\bibitem [{\citenamefont {Wei}\ \emph {et~al.}(2007)\citenamefont {Wei},
  \citenamefont {Sharma}, \citenamefont {Nunez}, \citenamefont {Haney},
  \citenamefont {Duine}, \citenamefont {Bass}, \citenamefont {MacDonald},\ and\
  \citenamefont {Tsoi}}]{tsoi2007prl}%
  \BibitemOpen
  \bibfield  {author} {\bibinfo {author} {\bibfnamefont {Z.}~\bibnamefont
  {Wei}}, \bibinfo {author} {\bibfnamefont {A.}~\bibnamefont {Sharma}},
  \bibinfo {author} {\bibfnamefont {A.~S.}\ \bibnamefont {Nunez}}, \bibinfo
  {author} {\bibfnamefont {P.~M.}\ \bibnamefont {Haney}}, \bibinfo {author}
  {\bibfnamefont {R.~A.}\ \bibnamefont {Duine}}, \bibinfo {author}
  {\bibfnamefont {J.}~\bibnamefont {Bass}}, \bibinfo {author} {\bibfnamefont
  {A.~H.}\ \bibnamefont {MacDonald}}, \ and\ \bibinfo {author} {\bibfnamefont
  {M.}~\bibnamefont {Tsoi}},\ }\Doi {10.1103/PhysRevLett.98.116603} {\bibfield
  {journal} {\bibinfo  {journal} {Phys. Rev. Lett.},\ }\textbf {\bibinfo
  {volume} {98}},\ \bibinfo {pages} {116603} (\bibinfo {year}
  {2007})}\BibitemShut {NoStop}%
\bibitem [{\citenamefont {{Nogu{\'e}s}}\ \emph {et~al.}(2005)\citenamefont
  {{Nogu{\'e}s}}, \citenamefont {{Sort}}, \citenamefont {{Langlais}},
  \citenamefont {{Skumryev}}, \citenamefont {{Suri{\~n}ach}}, \citenamefont
  {{Mu{\~n}oz}},\ and\ \citenamefont {{Bar{\'o}}}}]{nogues2005pr}%
  \BibitemOpen
  \bibfield  {author} {\bibinfo {author} {\bibfnamefont {J.}~\bibnamefont
  {{Nogu{\'e}s}}}, \bibinfo {author} {\bibfnamefont {J.}~\bibnamefont
  {{Sort}}}, \bibinfo {author} {\bibfnamefont {V.}~\bibnamefont {{Langlais}}},
  \bibinfo {author} {\bibfnamefont {V.}~\bibnamefont {{Skumryev}}}, \bibinfo
  {author} {\bibfnamefont {S.}~\bibnamefont {{Suri{\~n}ach}}}, \bibinfo
  {author} {\bibfnamefont {J.~S.}\ \bibnamefont {{Mu{\~n}oz}}}, \ and\ \bibinfo
  {author} {\bibfnamefont {M.~D.}\ \bibnamefont {{Bar{\'o}}}},\ }\Doi
  {10.1016/j.physrep.2005.08.004} {\bibfield  {journal} {\bibinfo  {journal}
  {Phys.\ Rep.},\ }\textbf {\bibinfo {volume} {422}},\ \bibinfo {pages} {65}
  (\bibinfo {year} {2005})}\BibitemShut {NoStop}%
\bibitem [{\citenamefont {{Ali}}\ \emph {et~al.}(2003)\citenamefont {{Ali}},
  \citenamefont {{Marrows}}, \citenamefont {{Al-Jawad}}, \citenamefont
  {{Hickey}}, \citenamefont {{Misra}}, \citenamefont {{Nowak}},\ and\
  \citenamefont {{Usadel}}}]{ali2003prb}%
  \BibitemOpen
  \bibfield  {author} {\bibinfo {author} {\bibfnamefont {M.}~\bibnamefont
  {{Ali}}}, \bibinfo {author} {\bibfnamefont {C.~H.}\ \bibnamefont
  {{Marrows}}}, \bibinfo {author} {\bibfnamefont {M.}~\bibnamefont
  {{Al-Jawad}}}, \bibinfo {author} {\bibfnamefont {B.~J.}\ \bibnamefont
  {{Hickey}}}, \bibinfo {author} {\bibfnamefont {A.}~\bibnamefont {{Misra}}},
  \bibinfo {author} {\bibfnamefont {U.}~\bibnamefont {{Nowak}}}, \ and\
  \bibinfo {author} {\bibfnamefont {K.~D.}\ \bibnamefont {{Usadel}}},\ }\Doi
  {10.1103/PhysRevB.68.214420} {\bibfield  {journal} {\bibinfo  {journal}
  {Phys.\ Rev.\ B},\ }\textbf {\bibinfo {volume} {68}},\ \bibinfo {pages}
  {214420} (\bibinfo {year} {2003})}\BibitemShut {NoStop}%
\bibitem [{\citenamefont {{Georges}}\ \emph {et~al.}(2009)\citenamefont
  {{Georges}}, \citenamefont {{Grollier}}, \citenamefont {{Cros}},
  \citenamefont {{Fert}}, \citenamefont {{Fukushima}}, \citenamefont
  {{Kubota}}, \citenamefont {{Yakushijin}}, \citenamefont {{Yuasa}},\ and\
  \citenamefont {{Ando}}}]{georges2009prb}%
  \BibitemOpen
  \bibfield  {author} {\bibinfo {author} {\bibfnamefont {B.}~\bibnamefont
  {{Georges}}}, \bibinfo {author} {\bibfnamefont {J.}~\bibnamefont
  {{Grollier}}}, \bibinfo {author} {\bibfnamefont {V.}~\bibnamefont {{Cros}}},
  \bibinfo {author} {\bibfnamefont {A.}~\bibnamefont {{Fert}}}, \bibinfo
  {author} {\bibfnamefont {A.}~\bibnamefont {{Fukushima}}}, \bibinfo {author}
  {\bibfnamefont {H.}~\bibnamefont {{Kubota}}}, \bibinfo {author}
  {\bibfnamefont {K.}~\bibnamefont {{Yakushijin}}}, \bibinfo {author}
  {\bibfnamefont {S.}~\bibnamefont {{Yuasa}}}, \ and\ \bibinfo {author}
  {\bibfnamefont {K.}~\bibnamefont {{Ando}}},\ }\Doi
  {10.1103/PhysRevB.80.060404} {\bibfield  {journal} {\bibinfo  {journal}
  {Phys.\ Rev.\ B},\ }\textbf {\bibinfo {volume} {80}},\ \bibinfo {pages}
  {060404} (\bibinfo {year} {2009})}\BibitemShut {NoStop}%
\bibitem [{\citenamefont {{Chshiev}}\ \emph {et~al.}(2008)\citenamefont
  {{Chshiev}}, \citenamefont {{Theodonis}}, \citenamefont {{Kalitsov}},
  \citenamefont {{Kioussis}},\ and\ \citenamefont
  {{Butler}}}]{chshiev2008ieeem}%
  \BibitemOpen
  \bibfield  {author} {\bibinfo {author} {\bibfnamefont {M.}~\bibnamefont
  {{Chshiev}}}, \bibinfo {author} {\bibfnamefont {I.}~\bibnamefont
  {{Theodonis}}}, \bibinfo {author} {\bibfnamefont {A.}~\bibnamefont
  {{Kalitsov}}}, \bibinfo {author} {\bibfnamefont {N.}~\bibnamefont
  {{Kioussis}}}, \ and\ \bibinfo {author} {\bibfnamefont {W.~H.}\ \bibnamefont
  {{Butler}}},\ }\Doi {10.1109/TMAG.2008.2002605} {\bibfield  {journal}
  {\bibinfo  {journal} {IEEE Trans. Magn.},\ }\textbf {\bibinfo {volume}
  {44}},\ \bibinfo {pages} {2543} (\bibinfo {year} {2008})}\BibitemShut
  {NoStop}%
\bibitem [{\citenamefont {{Manchon}}\ \emph {et~al.}(2008)\citenamefont
  {{Manchon}}, \citenamefont {{Ryzhanova}}, \citenamefont {{Vedyayev}},
  \citenamefont {{Chschiev}},\ and\ \citenamefont {{Dieny}}}]{manchon2008jpcm}%
  \BibitemOpen
  \bibfield  {author} {\bibinfo {author} {\bibfnamefont {A.}~\bibnamefont
  {{Manchon}}}, \bibinfo {author} {\bibfnamefont {N.}~\bibnamefont
  {{Ryzhanova}}}, \bibinfo {author} {\bibfnamefont {A.}~\bibnamefont
  {{Vedyayev}}}, \bibinfo {author} {\bibfnamefont {M.}~\bibnamefont
  {{Chschiev}}}, \ and\ \bibinfo {author} {\bibfnamefont {B.}~\bibnamefont
  {{Dieny}}},\ }\Doi {10.1088/0953-8984/20/14/145208} {\bibfield  {journal}
  {\bibinfo  {journal} {J.\ Phys.\ Condens.\ Matter.},\ }\textbf {\bibinfo
  {volume} {20}},\ \bibinfo {pages} {145208} (\bibinfo {year}
  {2008})}\BibitemShut {NoStop}%
\bibitem [{\citenamefont {{Dimitrov}}\ \emph {et~al.}(2009)\citenamefont
  {{Dimitrov}}, \citenamefont {{Gao}}, \citenamefont {{Wang}}, \citenamefont
  {{Jung}}, \citenamefont {{Lou}},\ and\ \citenamefont
  {{Heinonen}}}]{Dimitrov2009jap}%
  \BibitemOpen
  \bibfield  {author} {\bibinfo {author} {\bibfnamefont {D.~V.}\ \bibnamefont
  {{Dimitrov}}}, \bibinfo {author} {\bibfnamefont {Z.}~\bibnamefont {{Gao}}},
  \bibinfo {author} {\bibfnamefont {X.}~\bibnamefont {{Wang}}}, \bibinfo
  {author} {\bibfnamefont {W.}~\bibnamefont {{Jung}}}, \bibinfo {author}
  {\bibfnamefont {X.}~\bibnamefont {{Lou}}}, \ and\ \bibinfo {author}
  {\bibfnamefont {O.}~\bibnamefont {{Heinonen}}},\ }\Doi {10.1063/1.3137197}
  {\bibfield  {journal} {\bibinfo  {journal} {J.\ Appl.\ Phys.},\ }\textbf
  {\bibinfo {volume} {105}},\ \bibinfo {pages} {113905} (\bibinfo {year}
  {2009})}\BibitemShut {NoStop}%
\end{thebibliography}
\end{document}